\documentclass[prl,citeautoscript, reprint, amsmath, amssymb]{revtex4-2}

\usepackage{bm}
\usepackage{amsmath}
\usepackage{mathrsfs}
\usepackage{graphicx}
\usepackage{dcolumn}
\usepackage{bm}
\usepackage{lipsum}
\usepackage{gensymb}
\usepackage{esvect} 
\usepackage{balance}
\usepackage{float}
\usepackage{makecell}
\usepackage{booktabs}
\usepackage{natbib}

\usepackage[utf8]{inputenc}
\usepackage[english]{babel}

\usepackage{hyperref}
\hypersetup{ 
    colorlinks=true,
    linkcolor=blue,
    filecolor=magenta,      
    urlcolor=cyan,
    pdftitle={Pfeifer2026},
    pdfpagemode=FullScreen,
    citecolor=,
    }

\newcommand{\rpm}{\sbox0{$1$}\sbox2{$\scriptstyle\pm$}\raise\dimexpr(\ht0-\ht2)/2\relax\box2 }

\begin{document}

\preprint{JRN/123-ABC}

\title{Phonon selection rules and interference in momentum-resolved electron energy loss spectroscopy}

\author{Thomas W. Pfeifer}
\affiliation{Center for Nanophase Materials Sciences, Oak Ridge National Laboratory, Oak Ridge,  Tennessee 37830, USA}
\affiliation{previously Department of Mechanical and Aerospace Engineering, University of Virginia, Charlottesville, Virginia 22904, USA}

\author{Harrison A. Walker}
\affiliation{Interdisciplinary Materials Science Program, Vanderbilt University, Nashville, Tennessee 37235, USA}

\author{Henry Aller}
\affiliation{Department of Mechanical Engineering, University of Maryland, College Park, Maryland 20742, USA}

\author{Samuel Graham}
\affiliation{Department of Mechanical Engineering, University of Maryland, College Park, Maryland 20742, USA}

\author{Sokrates T. Pantelides}
\affiliation{Interdisciplinary Materials Science Program, Vanderbilt University, Nashville, Tennessee 37235, USA}
\affiliation{Department of Physics and Astronomy, Vanderbilt University, Nashville, Tennessee 37235, USA}

\author{Jordan A. Hachtel}
\affiliation{Center for Nanophase Materials Sciences, Oak Ridge National Laboratory, Oak Ridge,  Tennessee 37830, USA }

\author{Patrick E. Hopkins}
\email{peh4v@virginia.edu}
\affiliation{Department of Mechanical and Aerospace Engineering, University of Virginia, Charlottesville, Virginia 22904, USA}
\affiliation{Department of Physics, University of Virginia, Charlottesville, Virginia 22904, USA}
\affiliation{Department of Materials Science and Engineering, University of Virginia, Charlottesville, Virginia 22904, USA}

\author{Eric R. Hoglund}
\email{hoglunder@ornl.gov}
\affiliation{Center for Nanophase Materials Sciences, Oak Ridge National Laboratory, Oak Ridge,  Tennessee 37830, USA}


\date{\today}

\begin{abstract}
    As momentum-resolved Electron Energy Loss Spectroscopy (q-EELS) becomes more widely used for phonon measurements, better understanding of the intricacies of the acquired signal is necessary. Selection rules limit the allowed scattering, which may prohibit the appearance of specific phonon branches in some measurements. Simultaneous sampling of the lattice across all basis atoms also warrants a coherent treatment of phonons, which yields a reciprocal-space repeating unit cell that is larger than the standard Brillouin zone. We thus introduce the concept of the ``interferometric Brillouin zone'' where phonons are observed, which is closely related to the Dynamic Structure Factor. These effects follow from our new mathematical formalism for waves and vibrations, and we demonstrate calculations of q-EELS experiments using molecular dynamics and lattice dynamics. Results are compared to established q-EELS simulation methods in well-studied material systems, including the utilization of scattering selection rules to acquire a polarization-selective vibrational density of states. Finally, we note the analysis involved is directly applicable to any wave phenomena, such as plasmons or polaritons.
\end{abstract} 

\maketitle

\newpage

\section{Introduction}


Increased energy and momentum resolution in Electron Energy Loss Spectroscopy (EELS) has enabled measurements of phonon spectra in the electron microscope \cite{Senga2019, Nicholls2019, Lovejoy2020, Li2024}. These measurements are critical for a variety of applications. Thermal management in high-power applications and micro-electronics relies on fundamental understanding and characterization of phonon behavior. The ability to measure localized and polarization-dependent vibrational densities of states (v-DOS) also offers the opportunity to expand our understanding of phonon behavior as it relates to nano-structure, interfaces, and defects \cite{Hoglund2023,Hoglund2024,Yang2024,Li2024,Haas2024}. Finally, phonons with unique topological states are of interest for a range of quantum applications including sensing and computing. 

In order to adequately interpret phonon dispersions and behavior measured via vibrational momentum-resolved EELS (q-EELS), a thorough understanding of phonon interference and of the electron beam scattering physics is necessary. Recent works have developed computational tools for simulating q-EELS using molecular dynamics and electron multislice calculations \cite{Zeiger2021,CR2025}. Others have developed the theoretical framework based on the passing electron's excitation of the crystal's vibrational states \cite{Nicholls2019,Senga2019,Rossi2024}.

In this work, we construct a relatively simpler formulation of the problem within a MD framework by calculating a coherent probability density for the wavelike vibrational motion of atoms. Wave interference between basis sites within the crystal results in silencing of select phonon modes in certain Brillouin zones. We thus introduce the concept of the ``interferometric Brillouin zone'', which is a new minimum repeating unit in reciprocal space, which may encompass more than one conventional Brillouin zone. In other words, when destructive interference is considered, the relative motion of atoms may not be identical within all Brillouin zones. 

We also highlight the direction selectivity rules which govern the scattering of fast-electrons by phonons. The degree of scattering of the electron depends in part on the phonon eigenvector direction, which further affects which branches may appear in a q-EELS experiment. These phenomena exist in both simulations and experiments, and exist separately from issues of instrument resolution or sensitivity. These concepts will be discussed in this work within the context of phonon EELS, however they are also applicable to other spectroscopic experiments and the measurement of other quasi-particles. 

Our formulation enables the evaluation of the effects which are linked to the fundamental behavior of phonons (i.e. the destructive interference effects, which can merely be reproduced with a change of basis) v.s. those which are the direct result of the scattering of electrons. Our methodology also serves as a tool for replicating the kinematic scattering behavior in q-EELS at a fraction of the computational cost of conventional electron multislice. Results are fully consistent with the dynamical structure factor that underlies treatment of neutron and X-ray scattering from liquids, amorphous materials, and crystalline solids \cite{Dorner1982, Burkel2001,Sinha2001}, and are compared to spectral energy density calculations \cite{Dove1993,Thomas2010} conventionally used to calculate the phonon dispersion from MD. 

 Finally, we include a number of key observations based on simulations. We demonstrate the principle of acquiring a polarization-selective vDOS via dark-field EELS, and we note the degree of selectivity depends on the STEM probe convergence-angle. We also note the presence of thickness effects, where the simulations (and prospective measurements) show heightened sensitivity to the upper surface of the sample. 

All details on molecular dynamics (performed in LAMMPS \cite{LAMMPS}), TACAW \cite{CR2025} and FRFPMS \cite{Zeiger2021} simulations, multislice (using abTEM \cite{abTEM}), and LD (using phonopy \cite{phonopy1,phonopy2}) are available in the appendices and Supplemental Material.

\section{Theoretical background of wave coherence}
The coherent vs. incoherent behavior or phonons is alluded to in the works of Li \textit{et al.} \cite{Li2024}, however we will explore this rigorously here. Phonons are quantized lattice waves \cite{Chen2005,Kittel2005,Srivastava2022}, i.e., wavelike vibrations in crystals. Each mode is described by its frequency $\omega$, wavevector $\vv{\bm{k}}$ (with the magnitude being the inverse of the wavelength), and number of quanta (or population, akin to the wave amplitude in the classical picture). We begin by considering waves generally however, as the concepts discussed are closely related to interference phenomena observed for other quantum particles such as electrons or photons. 


When considering any quantum-mechanical system containing two or more waves (e.g. $\Psi_1$,$\Psi_2$,...), opposing phases can yield partial or total destructive interference. This phenomena is referred to as ``coherent superposition''. When considering the $\Psi$ of the combined system, component waves are summed: $\Psi_{system}=\Sigma_{j}\Psi_j$, and the probability density ($\rho$) is defined as the square (Eq. \ref{eq:rhoco}, note the star denoting the complex conjugate). This coherent probability density includes interference or ``cancellation'' between different waves. Alternatively, the system can be evaluated incoherently, i.e., the behavior of individual waves is considered independently, and no destructive interference is captured. The incoherent probability density is defined as the sum of individual probability densities (Eq. \ref{eq:rhoinco}), i.e., the magnitude is taken prior to summation. 

\begin{equation}
\rho_{coherent}=|\Psi_{system}|^2=|\Sigma_{j}\Psi_j|^2=\Psi_{system}^* \Psi_{system}
\label{eq:rhoco}
\end{equation}

\begin{equation}
\rho_{incoherent}=\sum_j|\Psi_j|^2=\sum_j\Psi_j^*\Psi_j
\label{eq:rhoinco}
\end{equation}

Note that summation and interference effects can occur in either real or reciprocal space, assuming the summed waves have the same frequency and wavelength. For two $\vv{\bm{\Psi}}$ at differing $\vv{\bm{k}}$ or $\omega$ however, no interference occurs, as the waves vary between being in- and out-of-phase (a "beating" phenomena with a net-zero change in wave magnitude over all of time and space). This can be shown mathematically via the fourier transform for any periodic signal: $\mathscr{F}[e^{i \, \omega_l \, t}]=\delta(\omega-\omega_l)$, where oscillations at differing frequencies $\omega_l$ and $\omega_l'$: $\delta(\omega-\omega_l)$ and $\delta(\omega-\omega_l')$, do not interact elastically. 


A generalized expression for $\Psi$ at a given wavevector $\vv{\bm{k}}$ and frequency $\omega$ may take the form:
\begin{equation}
	\vv{\bm{\Psi}}_{\vv{\bm{k}},\omega}(\vv{\bm{x}},t)= \vv{\bm{\Lambda}}_{\vv{\bm{k}},\omega} \; e^{i \, (\vv{\bm{k}} \bullet \vv{\bm{x}} - \omega \, t)}
	\label{eq:psigeneral}
\end{equation}
which is simply a plane wave varying in time and along position vector $\vv{\bm{x}}$. The amplitude, polarization, and phase of the wave are captured by the 3D complex vector $\vv{\bm{\Lambda}}$. Note that in many cases, specific terms are excluded, combined, or vectors are treated as scalars (e.g. for 1D or where polarization is not required), however we have elected to present a more general form above.  

In the case of phonons, waves are comprised of oscillatory atomic motion occurring across the system. The wave is sampled at discrete points in space (i.e., at lattice sites), meaning the real-space atomic position vector $\vv{\bm{r}}_{n,j}$ for atom $j$ in unit cell $n$ is used in place of $\vv{\bm{x}}$. At a single wavevector, multiple branches may also exist, so we introduce indices $l$ and $m$ to index a mode at $\vv{\bm{k}}_l$,$\omega_m$. The polarization and phase are often represented by unity-magnitude eigenvector $\vv{\bm{\varepsilon}}$, and the amplitude (in the classical case) is scalar $A$, i.e., $\vv{\bm{\Lambda}}$ from Eq. \ref{eq:psigeneral} is  $\vv{\bm{\Lambda}}=A \, \vv{\bm{\varepsilon}}$. The polarization, direction, and magnitude may differ between modes ($l$,$m$) and across the basis ($j$), hence the inclusion of these additional indices in the mode-wise eigenvector $\vv{\bm{\varepsilon}}_{j,l,m}$ and mode-wise amplitude $A_{j,l,m}$ below. We note that differing conventions exist \cite{Srivastava2022} and some works use the lattice vector $\vv{\bm{r}}_{n}$ \cite{Kaxiras2003} while others use the position vector $\vv{\bm{r}}_{n,j}$ \cite{Kresse1995,phonopy1,phonopy2}. A difference in phase for the resulting $\vv{\bm{\varepsilon}}$ will occur, according to $e^{i \,  \vv{\bm{k}}\bullet\vv{\bm{s}}_j}$, if the atom's relative position within the unit cell $\vv{\bm{s}}_j$ is defined by $\vv{\bm{r}}_{n,j}=\vv{\bm{r}}_{n,j=0}+\vv{\bm{s}}_{j}$ (additional details available in the Supplemental Material). For the purposes of this paper, we choose the $\vv{\bm{r}}_{n,j}$ convention, meaning the vibrations of a single atom (in unit cell $n$ at basis index $j$) due to a specific phonon (indexed by $l$,$m$) are thus \cite{Srivastava2022}: 
\begin{equation}
	\vv{\bm{u}}_{n,j,l,m}(t)=
	A_{j,l,m} \; \vv{\bm{\varepsilon}}_{j,l,m} \; e^{ i \: ( \vv{\bm{k}}_l \bullet \vv{\bm{r}}_{n,j} -  \omega_m \, t)}
	\label{eq:umode}
\end{equation}

We highlight the fact that a different displacement amplitude ($A_j$) and eigenvector ($\vv{\bm{\varepsilon}}_j$) are calculated for each basis index ($j$), for a given value of $\omega$ and $\vv{\bm{k}}$. In other words, multiple waves coexist within the system for a single mode. The multiple waves can be seen in numerous derivations \cite{Chen2005,Kittel2005,Srivastava2022}, most commonly with a simplified 1D diatomic chain model (``Born von Karman model'' of lattice dynamics (LD)); two eigenvalues are calculated to determine the amplitude of oscillations, and the sign of these eigenvalues determines the in- or out-of-phase motion of the masses. This is seen in conventional LD packages in 3D as well; 3$B$ complex eigenvalues are calculated for $B$ atoms within the basis for each $\omega_m$ and $\vv{\bm{k}}_l$ \cite{phonopy1,phonopy2}. 

In molecular dynamics (MD), the mode-wise displacements $\vv{\bm{u}}_{n,j,l,m}(t)$ are not immediately accessible, and vibrations may exist which are not strictly phonons (e.g. wave-like propagons in amorphous materials, or short-lived diffusons \cite{Allen1999}). The time-dependent motion of the individual atoms in a thermalized system is the sum over all vibrational modes, plus anharmonic contributions:

\begin{equation}
	\vv{\bm{u}}_{n,j}(t)=\sum_{l,m} \vv{\bm{u}}_{n,j,l,m}(t) + \vv{\bm{u}}_{n,j}^{anharmonic}
	\label{eq:utotal}
\end{equation}

The convention we adopted for the above equations, using atomic position vector $\vv{\bm{r}}_{n,j}$ as opposed to unit cell position $\vv{\bm{r}}_{n}$, is critical, as it allows consistent treatment of arbitrarily complex structures. Within MD, the displacements $\vv{\bm{u}}_{n,j}(t)$ can be defined in terms of the average position of each atom $n,j$. Supercells can be arbitrarily complex, and amorphous materials can be analyzed by taking all atoms in the supercell to be a single basis of a supercrystal.

While $\vv{\bm{u}}_{n,j}(t)$ can be directly recorded in MD, our objective is to unpack $\vv{\bm{u}}_{n,j}(t)$ into $\vv{\bm{k}}$,$\omega$ space phonon modes, i.e., to recover the displacement amplitudes of across all basis indices $j$ for each mode $l$,$m$. 
To this end, we perform a spatio-temporal Fourier transform of $\vv{\bm{u}}_{n,j}(t)$. This transform is discrete in space: $X_k=\frac{1}{N}\sum x_n \; e^{-i \, k \, n}$, and continuous in time: $\mathscr{F}(\omega)=\frac{1}{\tau_f}\int_{0}^{\tau_f} f(t) \; e^{i \, \omega \, t} dt$:
\begin{equation} \begin{split}
		\vv{\bm{\Psi}}_{j}(\vv{\bm{k}},\omega) =
		\vv{\bm{u}}_j(\vv{\bm{k}},\omega) = 
		\;\;\;\;\;\;\;\;\;\;\;\;\;\;\;\;\;\;\;\;\;\;\;\;\;\;\;\;\;\;\;\;\;\;\;\;\;\; \\
		\frac{1}{N \, \tau_f}
		\sum_{n}^{N} \int_{0}^{\tau_f} \vv{\bm{u}}_{n,j}(t) \;
		e^{- i \, ( \vv{\bm{k}} \bullet \vv{\bm{r}}_{n,j} - \omega \, t)} dt 
		\label{eq:psiphonon}
\end{split}
\end{equation}
where $\tau_{f}$ is the duration of the MD simulation. Complex 3D vector $\vv{\bm{u}}_j(\vv{\bm{k}},\omega)$ captures the wavelike vibrations occurring within the system at arbitrary $\vv{\bm{k}}$,$\omega$, which captures finite-temperature and anharmonic effects in addition to phonons. 

By plugging Eq. \ref{eq:psiphonon} into Eq. \ref{eq:rhoco} and \ref{eq:rhoinco} respectively, we find the expressions:

\begin{equation} \label{eq:SEDco} 
	\vv{\bm{\rho}}_{coh}(\omega,\vv{\bm{k}})
	=
	\frac{1}{N \, \tau_f}
	\bigg|
	\int_{0}^{\tau_{f}} 
	\sum_{j}^{B}
	\sum_{n}^{N_u}
	\vv{\bm{u}}_{n,j}(t) \;
	e^{ - i \, ( \vv{\bm{k}} \bullet \vv{\bm{r}}_{n,j} - \omega \, t ) } dt
	\bigg|^{2}
\end{equation}

\begin{equation} \label{eq:SEDinco} 
	\vv{\bm{\rho}}_{inco}(\omega,\vv{\bm{k}})
	=
	\frac{1}{N \, \tau_f}
	\sum_{j}^{B}
	\bigg|
	\int_{0}^{\tau_{f}} 
	\sum_{n}^{N_u}
	\vv{\bm{u}}_{n,j}(t) \;
	e^{ - i \, ( \vv{\bm{k}} \bullet \vv{\bm{r}}_{n,j} - \omega \, t ) } dt
	\bigg|^{2}
\end{equation}

$\vv{\bm{\rho}}_{inco}(\omega,\vv{\bm{k}})$ thus represents the phonon density within the system under the traditional viewpoint: eigendisplacements associated with each crystal basis index are treated independently, and destructive interference between optical modes (where atoms in the basis vibrate out of phase) is not considered. Conversely, $\vv{\bm{\rho}}_{coh}(\omega,\vv{\bm{k}})$ represents the net wavelike vibrations with quantum superposition included, i.e., destructive interference may occur.

A similar coherent calculation can be performed for LD, despite the wave amplitudes not being calculated. Atomic displacements are also not found in LD, as the eigenvectors $\vv{\bm{\varepsilon}}_j$ are calculated directly. If $\vv{\bm{u}}_{n,j}^{anharmonic}$ from Eq. \ref{eq:utotal} is zero, then we can separate the amplitude and eigenvector terms from Eq. \ref{eq:psiphonon}: $\vv{\bm{\Psi}}_j(\vv{\bm{k}},\omega) = \vv{\bm{u}}_j(\vv{\bm{k}},\omega) = A_j(\vv{\bm{k}},\omega) \, \vv{\bm{\varepsilon}}_j(\vv{\bm{k}},\omega)$. LD calculations are also performed at discrete $\vv{\bm{k}}_l$,$\omega_m$ as seen in Eq. \ref{eq:umode}. Therefore, neglecting the amplitude term and anharmonic contributions, the coherent and incoherent probability density can be calculated from LD by similarly plugging into Eq. \ref{eq:rhoco} and \ref{eq:rhoinco}, which yields:  


\begin{equation}
	\vv{\bm{\rho}}_{l,m,coh}=
	\bigg|
	\sum_j^B\vv{\bm{\varepsilon}}_{j,l,m}
	\bigg|^2
	\label{eq:LDcohe}
\end{equation}

\begin{equation}
	\vv{\bm{\rho}}_{l,m,inco}=
	\sum_j^B
	\bigg|
	\vv{\bm{\varepsilon}}_{j,l,m}
	\bigg|^2
	\label{eq:LDinco}
\end{equation}

\begin{figure}
	\centering
	\includegraphics[width=1\linewidth]{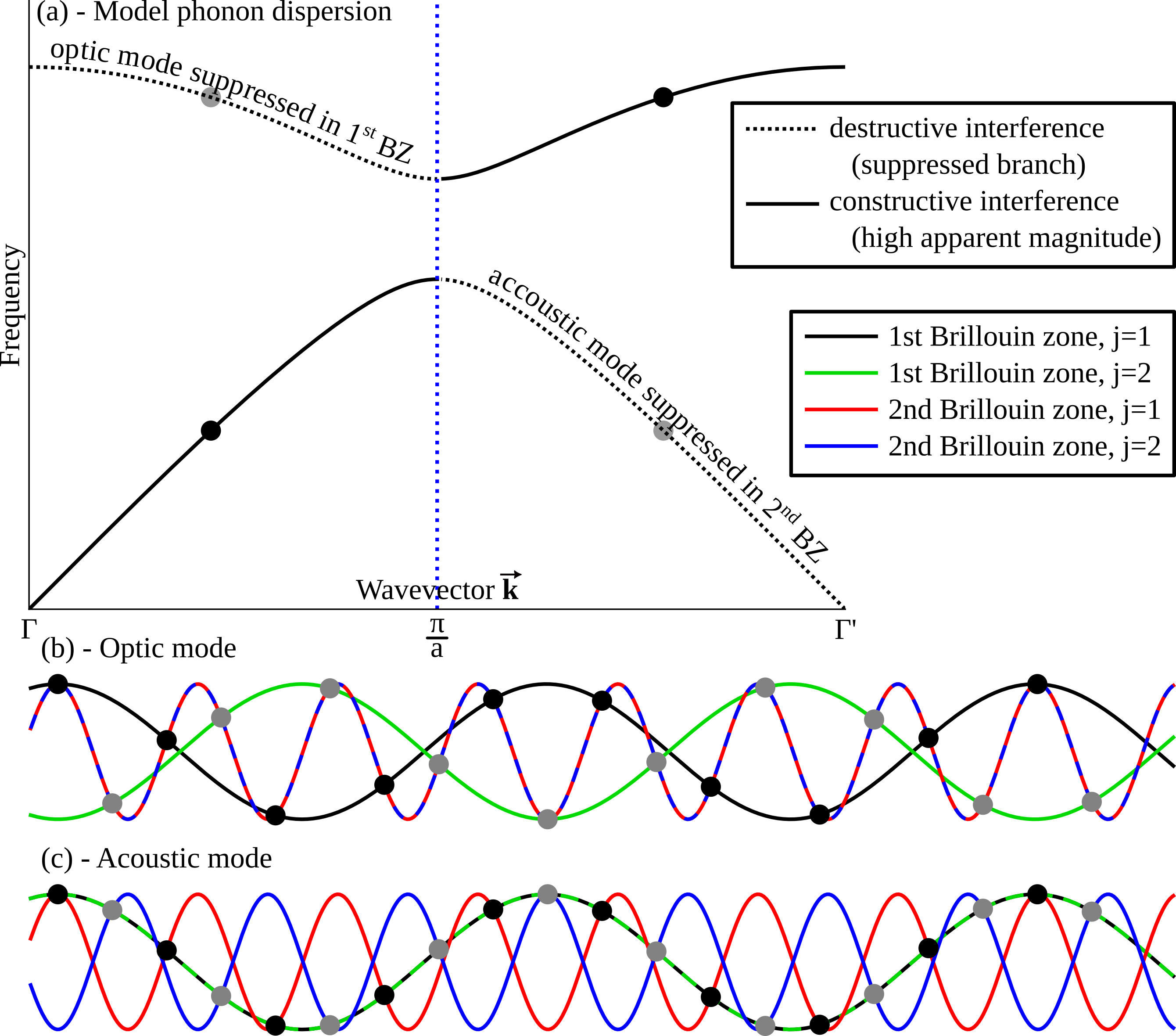}
	\caption{(a) A simplified diatomic phonon dispersion consists of an acoustic and optic branch. One optic and acoustic mode have been selected, along with the corresponding modes in the second Brillouin zone (yielding identical displacements). Displaced atoms (black and grey) are shown for optic and acoustic modes in (b) and (c) respectively, with the waves passing through each subset of atoms shown. For the optic mode (b), $\Psi_1$ (black) and $\Psi_2$ (green) are out of phase in the first Brillouin zone (small $|\protect\vv{\bm{k}}|$ or long wavelength). Within the next Brillouin zone however (large $|\protect\vv{\bm{k}}|$ or short wavelength), $\Psi_1$ (red) and $\Psi_2$ (blue) are in phase / overlapped. The opposite is true for the acoustic mode (c): $\Psi_1$ (black) and $\Psi_2$ (green) are in phase in the first Brillouin zone, but out of phase in the next (red and blue). The same phonon within the first and second Brillouin zones are thus non-equivalent, and a scattering experiment sensitive to both atomic basis sites will show differing sensitivity to the phonon as a result.}
	\label{fig:opticalphonon}
\end{figure}

A simplified depiction of the multiple waves is shown in Figure \ref{fig:opticalphonon} for both optical and acoustic phonons. For the optical phonon (Fig. \ref{fig:opticalphonon}.b), the atoms are traditionally discussed as vibrating out of phase (i.e., the signs of $\vv{\bm{\varepsilon}}_{j=1}$ and $\vv{\bm{\varepsilon}}_{j=2}$ are opposite in the simplest diatomic case). This may not be the case in higher-order Brillouin zones however. Within the first Brillouin zone (small $|\vv{\bm{k}}|$, long wavelength, black and green lines in Fig. \ref{fig:opticalphonon}.b), the waves describing alternating atoms are equal and out of phase with respect to each other. Within the second Brillouin zone however (large $|\vv{\bm{k}}|$, short wavelength, blue and red in Fig. \ref{fig:opticalphonon}.b), the same atomic displacements are described by a sinusoid passing through both sets of atoms. In this manner, the mode in the 1\textsuperscript{st} and 2\textsuperscript{nd} Brillouin zones should not be considered equivalent, despite identical atomic displacements, once the phase agreement or cancellation of the two component waves is considered. 
A similar phenomena occurs for the acoustic mode (Fig. \ref{fig:opticalphonon}.c); within the second Brillouin zone, the two component waves (red and blue) are out of phase and destructive interference will occur. This concept is also very similar to the unfolding of super-cells based on crystallographic symmetry as opposed to compositional \cite{Zheng2016,Ikeda2017}, ignoring vacancies or substitutional atoms when considering crystal symmetry \cite{Abeles1963,Bellaiche2000,Beechem2007}, or the extended zone scheme for phonons in superlattices \cite{Simkin2000}. We will demonstrate the consequences of the coherent treatment within realistic systems in the following sections. 

Our $\vv{\bm{\rho}}_{inco}(\omega,\vv{\bm{k}})$ expression is nearly identical, with the exception of scaling and atomic mass terms, to the expression for Spectral Energy Density \cite{Dove1993,Thomas2010} (comparison available in the Supplemental Material). Their derivation seeks a careful quantification of mode-wise energy within the system however (noting that vibrational atomic displacements contain kinetic energy from motion, and potential energy from the deformation of bonds). By framing our expression in terms of a mere spatio-temporal Fourier transform over the displacements resulting from all waves within the system, we allow application of these concepts to other quasiparticles and fields. For any other wave-like field (e.g. modulation of charge density), the relevant magnitude or polarization is plugged in for $\Lambda$ or $\vv{\bm{\varepsilon}}$ in Eq. \ref{eq:psigeneral}. The instantaneous localized charge, should many waves exist, is the sum, as in Eq. \ref{eq:utotal}. One can then return to $\vv{\bm{k}}$,$\omega$ space via the spatio-temporal Fourier transform. If the field is continuous in space (as opposed to discretely sampled, as with phonons), the continuous spatial transform can be used instead. This would be similar to the density operator in traditional scattering physics, e.g. $\vv{\bm{\rho}}(\vv{\bm{q}},\omega)=\int
\sum_j b_j e^{i \, \vv{\bm{q}}\bullet\vv{\bm{r}}_j(t)} \; e^{-i \, \omega \, t} dt$ for net scattering amplitude $b_j$ for atom $j$ \cite{Dove1993}.

\section{phonon coherent probability density in MD}

\begin{figure*}[t] 
	\centering
	\includegraphics[width=.95\linewidth]{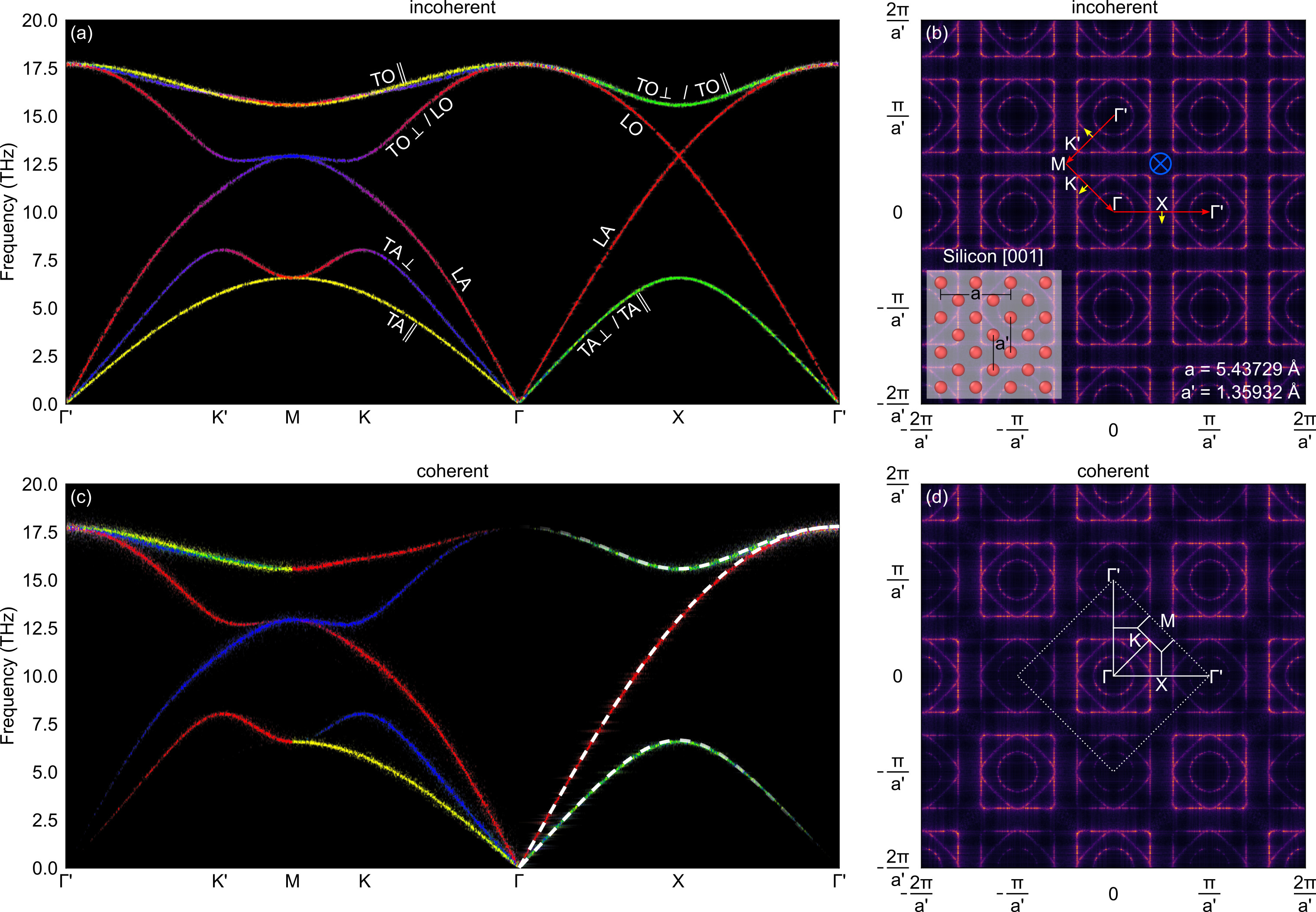}
	\caption{probability density calculations are performed for silicon (diatomic basis) using either incoherent (a,b), or coherent summing (c,d). Results are shown as a dispersion (a,c), with color used to denote eigenvector polarization (red, yellow, and blue denote eigenvector magnitudes along or perpendicular to the path, or through plane, respectively. this roughly corresponding to longitudinal, transverse in-plane, and transverse through-plane modes). RYB color mixing denotes degeneracy or mixed-polarization branches. Coherent LD is shown in dotted white in the $\Gamma$-$X$ direction in panel c. Iso-energy slices in reciprocal space are shown at 6 THz (b,d). In the coherent case, Brillouin zones are no longer identical, and we refer to the new larger minimum repeating unit in reciprocal space as an ``interferometric Brillouin zone''. The interferometric Brillouin zone is shown in dotted white in (d), and its size depends on the interatomic spacing as opposed to the size of the primitive cell. We also differentiate between non-equivalent $\Gamma$ and $K$ points with the $\Gamma'$ and $K'$ notation.}
	\label{fig:coherence}
\end{figure*}

Turning next to realistic molecular dynamics simulations, the coherent and incoherent probability density is compared in Figure \ref{fig:coherence}. In the incoherent case, points equidistant from the Brillouin zone edge are considered equivalent. At points beyond the first traditional Brillouin zone edge, the wavelength is smaller than the primitive cell. Each wave is sampled once per primitive cell prior to summation, meaning the duplication of the dispersion into outer Brillouin zones is merely the result of aliasing. In contrast, the coherent treatment yields an ``unfolding'' of optic modes across the Brillouin zone boundary, and acoustic modes may not appear within the second Brillouin zone. This is due to the interference effects discussed in Figure \ref{fig:opticalphonon}, where the phase agreement between waves on each sublattice yields constructive or destructive interference. Considering the integral in Eq. \ref{eq:SEDco} and \ref{eq:SEDinco} as integral transforms, with the plane wave as the kernel function, $\rho$ is thus the magnitude of the plane waves which match the displacements of atoms. The summation of across both $j$ and $n$ Eq. \ref{eq:SEDco} for $\rho_{coh}$ can therefore be thought of as describing a single wave sampled at each atomic site (as opposed to once per primitive cell). The minimum wavelength before aliasing is therefore smaller, and the minimum repeating area in reciprocal space is larger. We will refer to this larger area as the ``interferometric Brillouin zone''.

The phonon dispersions for Stillinger-Weber silicon are shown in Figure \ref{fig:coherence} in the [001] plane (i.e. the $\Gamma$-$X$ and $\Gamma$-$K$-$M$ directions). Longitudinal (L) vs. transverse (T) modes are differentiated based on the direction of the velocity and position vectors used for the calculation. For the incoherent case (considering atoms on each lattice site independently, Fig. \ref{fig:coherence}.a), we show the expected phonon dispersion in all directions. In the coherent case (sampling every atom simultaneously, \ref{fig:coherence}.c), the LO branch in [100] ($\Gamma$-$X$) is only present in the second Brillouin zone (between $\frac{2 \pi}{a}$ and $\frac{4 \pi}{a}$) as these vibrations are equal in magnitude and 180\degree{} out of phase for the two atoms in the basis. This means there is total destructive interference in the first Brillouin zone for the LO mode. TA and TO branches fade in and fade out, as there is partial and varying levels of interference. This is seen in calculations from both MD and LD (dashed white in Fig. \ref{fig:coherence}.c). In the [110] direction, similar behavior is observed for the previously-degenerate LA and TA\textsubscript{$\perp$} branches, where some branches may appear closer to the outer $\Gamma$ points. While Li \textit{et al.} \cite{Li2024} observed the systematic absence of phonon branches in q-EELS simulations, we would like to highlight that this is a fundamental behavior of phonons, and is observable even without the electron scattering effects included. 

We next considering probability density calculations across a grid of $k_x$,$k_y$ points at a fixed frequency, which is analogous to a hypothetical energy-resolved diffraction pattern in experiment. We have shown results at 6 THz in Fig. \ref{fig:coherence}.b,d, where the interferometric Brillouin zone is visible in the coherent case. We also see a non-equivalence for some $\Gamma$ points and $K$ points, and we adopt the ``gamma prime'' ($\Gamma'$) and ``K prime'' ($K'$) terminology. In the case of silicon, the interferometric Brillouin zone extends to $\Gamma'$, however in other materials the interferometric Brillouin zone may include one or more $\Gamma'$ points (examples for AlN in [001] and [010] planes are available in Supplemental Material). 

The size of the interferometric Brillouin zone is defined by the minimum interatomic spacing in a given direction (where we introduce the $a'$ vs. $a$ notation and so on, where $a'$ denotes the interatomic spacing and $a$ denotes the unit cell). This is because the minimum interatomic spacing (rather than the primitive cell size) defines the minimum sampling of waves within the system. In the virtual crystal approximation \cite{Abeles1963,Bellaiche2000,Beechem2007}, vacancy sites or substitutional atoms are not considered to break the crystallinity, and super-cells can be ``unfolded'' based on crystallographic symmetry \cite{Ikeda2017}. The concept of the interferometric Brillouin zone is thus very similar; coherent sampling is not sensitive to missing atoms, but \textit{is} sensitive to sampling on a periodicity smaller than the primitive unit cell. In the case of silicon, the interferometric Brillouin zone's real-space volume corresponds to a cube drawn around a single atom. The missing atoms on alternating tetrahedral sites do not affect the sampling of the wave, and filling in the missing atoms reduces the structure to simple cubic. 

The interferometric Brillouin zone can also be found in existing scattering physics; $\Gamma'$ corresponds to the forbidden Bragg diffraction points, while $\Gamma$ corresponds to the allowed points. The concept of the static structure factor and the interferometric Brillouin zone (tied to the Dynamic Structure Factor) are thus closely related. We believe the interferometric Brillouin zone concept is useful for understanding the phonon behavior universally however.

\begin{figure*}[t]
    \centering
    \includegraphics[width=0.95\linewidth]{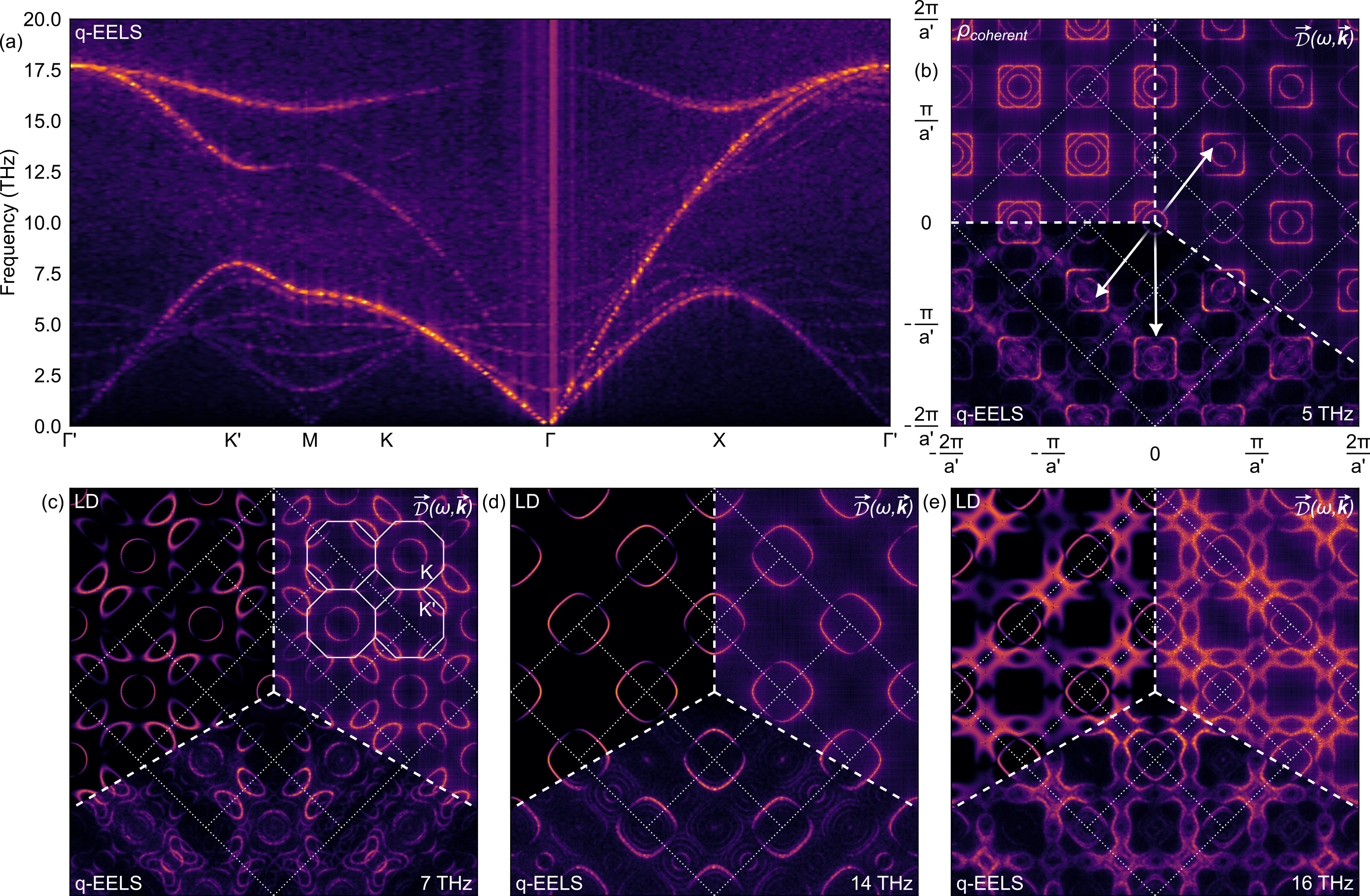}
    \caption{(a) phonon dispersions can be generated from q-EELS by collecting energy spectra along a reciprocal space path. Here the sampled path is identical to that shown in \ref{fig:coherence}.b, but centered on the [220] $\Gamma$ point so as to avoid suppression of transverse branches. Direction selectivity according to $\protect\vv{\bm{q}}\bullet\protect\vv{\bm{\varepsilon}}$ implies only phonons with eigenvectors $\protect\vv{\bm{\varepsilon}}$ in the direction of the electron scattering vector $\protect\vv{\bm{q}}$ will be visible. This is shown for the 5 THz energy slice (b) where coherent SED (without the application of selection rules) fails to replicate q-EELS. Upon the application of $\protect\vv{\bm{q}}\bullet\protect\vv{\bm{\varepsilon}}$ in SED however, the primary features from q-EELS can be replicated. Purely longitudinal and purely transverse branches appear as crescents, since the intensity fades to zero where the direction of atomic displacements $\protect\vv{\bm{\varepsilon}}$ are perpendicular to $\protect\vv{\bm{q}}$. There is also a near-complete suppression of modes comprised soley of through-plane vibrations (e.g. TO$_\perp$, blue branches from Fig. \ref{fig:coherence}). Additional energy-resolved diffraction images are shown for 7 THz (c), 14 THz (d) and 16 THz (e), generated via q-EELS, SED, and LD. The interferometric Brillouin zone is shown in dotted white, and the traditional Brillouin zone is shown in (c) in solid white. ``Unfolding'' behavior is also clearly visible: optical branches form crescents or circles about $\Gamma'$ at high frequencies, and there are ellipses about $K'$ points (and not $K$ points) at 7 THz.}
    \label{fig:dispersions}
\end{figure*}

\section{Selection rules}

The Dynamic Structure Factor (DSF) is commonly seen in inelastic electron, X-ray, and neutron scattering experiments \cite{Sturm1993,Burkel2001,Sinha2001}. DSF is nearly identical to our Eq. \ref{eq:SEDco} with the key exception of a atomic form factors, Debye-Waller scaling, and $\vv{\bm{q}} \bullet \vv{\bm{\varepsilon}}$ term instead of our $\vv{\bm{u}}$. A full derivation is available in the Supplemental Material, however we note the derivation involves a few key assumptions including: the Born approximation, harmonic vibrations in the crystal \cite{ashcroft1976}, and limited to single-phonon interactions. This should come as no surprise, as the Born approximation and Van Hove correlation function \cite{Hove1954} suggest that the scattering of particles (which do not meaningfully alter the scatterer populations) simply depends on the population that is present. For the single-phonon interaction case, scattering of the electron directly matches the population density, as multiple-scattering events are not considered.

Focusing on the $\vv{\bm{q}} \bullet \vv{\bm{\varepsilon}}$ term, this has been long-understood to result in polarization selectivity in INS and IXS experiments, with Ashcroft \& Mermin commenting on this in their appendix in 1976 \cite{ashcroft1976}. We feel the importance of this is greatly underappreciated in electron spectroscopy however. Practically, the dot product $\vv{\bm{q}} \bullet \vv{\bm{\varepsilon}}$ suggest that electrons only gains momentum in a given direction due to atomic displacements (or the component of atomic displacements) in that same direction. As noted by Nicholls \textit{et al.} \cite{Nicholls2019}, this means only certain vibrational modes will appear
in the spectra. This direction selectivity will result in the total absence of specific phonon branches in certain Brillouin zones, meaning caution is required in the interpretation of experimental or simulated results. We highlight that this direction selectivity is separate from the destructive interference effects discussed previously, and both will affect which branches are observed within a given Brillouin zone. Furthermore, this direction selectivity can yield an experimental sensitivity to phonon polarization if taken advantage of properly \cite{Hoglund2024}.


\renewcommand{\arraystretch}{1.5}
\begin{table*}[!htbp]
\begin{tabular}{lllll}
    Technique              &   FRFPMS   &   TACAW     &   $\vv{\bm{\rho}}(\omega,\vv{\bm{k}})_{coh}$ or $\vv{\bm{\mathcal{D}}}(\omega,\vv{\bm{k}})$   &   LD \\ \hline

    Description            & \makecell[lt]{ Frozen-phonon electron\\multislice calculations\\are performed from\\frequency-resolved (e.g.\\via a band-pass frequency\\filter) randomized\\snapshots from molecular\\dynamics. }  & \makecell[lt]{ Multislice calculations\\are performed over\\consecutive molecular\\dynamics snapshots and\\the resulting\\time-dependent exit wave\\is Fourier transformed. }  & \makecell[lt]{ Spatial and temporal\\Fourier analysis is\\performed over\\consecutive molecular\\dynamics snapshots. No\\multislice calculations\\are performed. }  & \makecell[lt]{ Harmonic vibrational\\motion is calculated from\\2nd order force constants\\alone. No molecular\\dynamics nor multislice\\is performed. } \\[75pt] \hline

    Advantages             & \makecell[lt]{ Full multislice captures\\kinematic and dynamic\\scattering physics.\\Randomized snapshots\\taken from band-pass\\filtered MD trajectory\\help ensure sufficient\\sampling over time }  & \makecell[lt]{ Full multislice captures\\kinematic and dynamic\\scattering physics.\\Consecutive snapshot and\\Fourier filter means\\fewer snapshots can be\\processed for a given\\energy resolution. }  & \makecell[lt]{ No multislice means\\results can be directly\\interpreted as phonon\\behavior. Coherent and\\incoherent analysis can\\be performed, and\\eigenvector selection can\\be enforced manually. No\\multislice calculation\\means lower computational\\cost. }  & \makecell[lt]{ Computation is performed\\without the need for\\molecular dynamics,\\meaning calculation is\\performed at\\significantly lower\\computational cost. } \\[105pt] \hline

    Disadvantages          & \makecell[lt]{ Finer energy resolution\\requires smaller\\band-pass filters and\\multislice is performed\\over significantly more\\snapshots. }  & \makecell[lt]{ Too few snapshots used\\may mean there is\\insufficient averaging\\across thermal\\fluctuations within the\\system. }  & \makecell[lt]{ Only kinemetic effects\\are captured, as no\\multislice is performed.\\Thickness-dependent\\electron scattering and\\multiple-scattering\\effects are not captured. }  & \makecell[lt]{ Traditional LD is \\ harmonic only. Phonon \\ linewidths and scattering \\ rates are not captured. } \\[64pt] \hline
\end{tabular}
\end{table*}

In Fig. \ref{fig:dispersions}, we show parallel-beam q-EELS simulations on the same silicon system shown previously. In Fig. \ref{fig:dispersions}.a, we present the energy spectrum along the same path in reciprocal space as in Fig. \ref{fig:coherence}.a,c, however an offset of 1 interferometric Brillouin zone has also been applied, since selection rules can prohibit the observation of some branches. The same branches are visible in their ``unfolded'' form, with the exception of those in the through-plane direction (blue in Fig. \ref{fig:coherence}), as displacements in [001] are parallel to the beam direction (i.e., no change in $q$ will be observed). 

In Fig. \ref{fig:dispersions}.b-e, we show several energy-resolved diffraction images, and additional energy levels are shown in the Supplemental Material. The bottom of each panel is generated via q-EELS simulations, and the other portions will be discussed later. At low-frequencies (Fig. \ref{fig:dispersions}.b,c), rings or crescents can be observed in the energy-resolved diffraction images, centered about each $\Gamma$ point, with a radius according to the phonon modes' wavevector $|\vv{\bm{k}}|$. At higher frequencies (Fig. \ref{fig:dispersions}.d,e), the unfolding of the phonon branches into the interferometric Brillouin zone again applies, with crescents converging towards each $\Gamma'$ point. The non-equivalence of $K$ and $K'$ points can also be seen in Fig. \ref{fig:dispersions}.c, with ellipses forming only about $K'$. 


Polarization selectivity effects due to the $\vv{\bm{q}}\bullet\vv{\bm{\varepsilon}}$ term in DSF are also visible. Within the first Brillouin zone, $\vv{\bm{q}}\bullet\vv{\bm{\varepsilon}}$ is only nonzero for modes where displacements are parallel to $\vv{\bm{q}}$, i.e., there is 100\% selectivity for longitudinal modes. This selectivity and the desire to show equivalent transverse modes between Figures \ref{fig:coherence} and \ref{fig:dispersions} thus motivated our use of a 1 interferometric Brillouin zone shift for generation of the dispersions. 

For outer interferometric Brillouin zones, we see crescents which fade to zero intensity where the direction of atomic displacements ($\vv{\bm{\varepsilon}}$) become perpendicular to the electron scattering direction ($\vv{\bm{q}}$). Longitudinal vs transverse modes can thus be differentiated based on the orientation of the crescent with respect to the center. Using 5 THz as an example (Fig. \ref{fig:dispersions}.b), the inner crescents correspond to LA modes, and a gap between crescents point away from the center. In contrast, the outer squares fade to zero intensity closest to the center, implying that this is a transverse branch with displacements perpendicular to the scattered electron. 

\section{Approximating q-EELS via the coherent probability density from MD or LD}

To replicate kinematic scattering and explore direction selectivity effects, we make a simplified modification to Eq. \ref{eq:SEDco} with the introduction of a $\vv{\bm{q}}\bullet\vv{\bm{\varepsilon}}$ term.  Recalling that our calculated $\vv{\bm{u}}_j(\vv{\bm{k}},\omega)$ term (Eq. \ref{eq:psiphonon}) contained eigenvector $\vv{\bm{\varepsilon}}$, the dynamic structure factor, neglecting scaling terms, is thus:

\begin{equation} \label{eq:DSF} 
	\vv{\bm{\mathcal{D}}}(\omega,\vv{\bm{k}})
	\propto
	\bigg|
	\int_{0}^{\tau_{f}} 
	\sum_{j}^{B}
	\sum_{n}^{N_u}
	\vv{\bm{q}} \bullet
	\vv{\bm{u}}_{n,j}(t) \;
	e^{ i \, ( \vv{\bm{k}} \bullet \vv{\bm{r}}_{n,j} - \omega \, t )} dt
	\bigg|^{2}
\end{equation}


This simplified approximation allows direct comparison with the results from Eq. \ref{eq:SEDco} and those in Fig. \ref{fig:coherence}.  Iso-energy slices from MD are shown in Fig. \ref{fig:dispersions}.c-e, upper right of each panel, using the expression above.

For calculations from LD, a similar treatment is applied to Eq. \ref{eq:LDcohe}:
 
 \begin{equation}\label{eq:DSFLD} 
 	\vv{\bm{\mathcal{D}}}(\omega,\vv{\bm{k}})
 	\propto
 	\bigg| \sum_j^B 
 	\vv{\bm{q}} \bullet
 	\vv{\bm{\varepsilon}}_j(\omega,\vv{\bm{k}})
 	\bigg|^2
\end{equation}
 
 Each complex 3D eigenvector $\vv{\bm{\varepsilon}}_j$ is calculated across a grid of $k$ points, coherent summation is applied, and the dot product is calculated. Whereas MD automatically includes thermal-broadening of phonon branches at finite temperature, no linewidths are typically calculated in LD. Iso-energy slices are thus generated by applying a gaussian linewidth to find the intensity as a function of $\vv{\bm{q}}$ at a given arbitrary frequency. LD results are shown in Figure \ref{fig:dispersions}.c-e, upper-left of each panel. 

There is reasonable agreement between the full rigorous multislice q-EELS simulations, LD, and $\vv{\bm{\mathcal{D}}}(\omega,\vv{\bm{k}})$ from MD. Differences and an in-depth comparison of the approximations involved will be discussed in the following section. 


\section{Comparing q-EELS vs the MD or LD approximations}

Thus far, we have shown the ``unfolding'' of the phonon Brillouin zone due to the coherent behavior of waves, and shown the effects of scattering selection rules. Both of these effects limit the appearance or reduce the intensity of branches on specific Brillouin zones, and are critical for replicating the primary features in the momentum-resolved EELS signal using LD or the coherent probability density ($\vv{\bm{\rho}}(\omega,\vv{\bm{k}})_{coh}$, Eq. \ref{eq:SEDco}) or our modified approximation for the dynamic structure factor ($\vv{\bm{\mathcal{D}}}(\omega,\vv{\bm{k}})$, Eq. \ref{eq:DSF}).

In this section we will discuss several key areas where $\vv{\bm{\mathcal{D}}}(\omega,\vv{\bm{k}})$ from MD or LD capture or fail to capture more subtle effects in the q-EELS signal. 

\begin{figure*}[t] 
    \centering
    \includegraphics[width=0.95\linewidth]{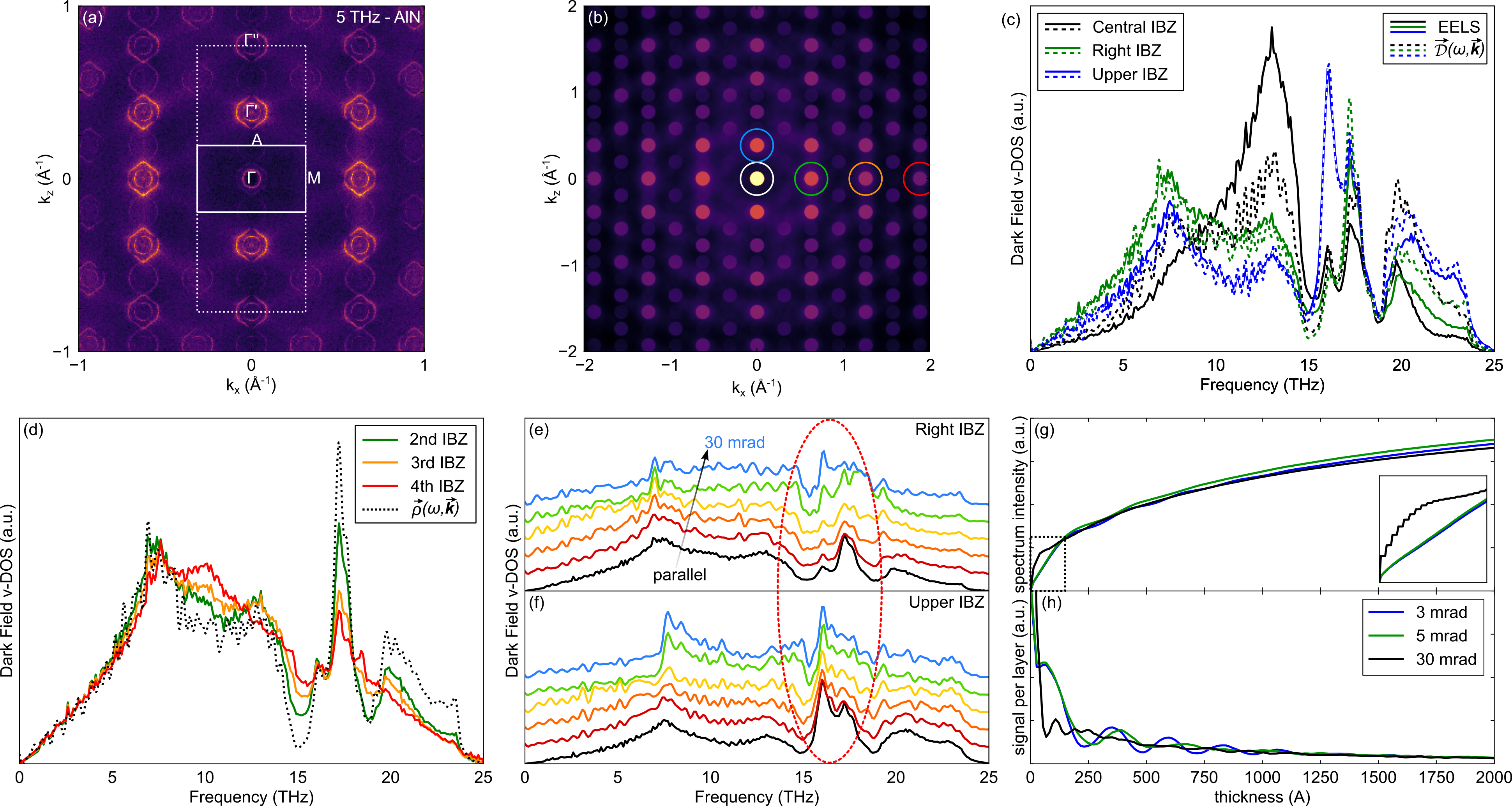}
    \caption{AlN in the [010] plane is used to examine polarization selection and the effects of a convergent beam. (a) an energy-resolved diffraction image is shown, which is used to inform selection of Brillouin zones and dark-field mask diameter. (b) circular masks are applied (with radius 1/$c$), centered on several $\Gamma$ points, shown here in the 3 mrad diffraction image. (c) We compare the DOS acquired from the central, right, and upper Brillouin zones (black, green, and blue, respectively) between parallel-beam q-EELS (solid) and DSF calculations from Eq. \ref{eq:DSF} (dashed). (d) Coherent probability density from Eq. \ref{eq:SEDco} yields a ``ground truth'' v-DOS (dashed), which we compare against the spectra acquired from the 2nd, 3rd, and 4th Brillouin zones. We see a greater deviation and therefore reduced polarization sensitivity at outer Brillouin zones, likely due to dynamic scattering effects. (e,f) parallel beam, and 1, 3, 5, 15, and 30 mrad convergence angles are used (black, red, orange, gold, green, blue), showing that direction selectivity is maintained for a convergent probe. Selectivity is reduced at large angles however. (g) The total q-EELS signal intensity is simulated as a function of depth (with finer resolution steps shown in the inset). (h) The signal on a per-layer basis is taken via the integral of (g). In both cases, the highest signal comes from the upper layers of the sample. Dynamic effects (Pendell{\"o}sung oscillations) can also be seen.}
    \label{fig:AlNDOS}
\end{figure*}

\subsection{Polarization-selective phonon density of states}

Direction selectivity resulted in the appearance of crescents in the energy-resolved diffraction images, however a dark-field spectrum acquisition may also be useful for measuring a polarization-selective vibrational density of states (v-DOS). We thus move to our second material system: AlN viewed in the [010] direction, which has a moderately anisotropic v-DOS between $\Gamma$-$K$-$M$ and $\Gamma$-$A$ directions. Under parallel beam illumination, the energy-resolved diffraction images appear as expected (example shown in Fig. \ref{fig:AlNDOS}.a), with the orientation of crescents identifying longitudinal and transverse modes.  We then apply a 1/$c$ radius mask in reciprocal space, centered on various $\Gamma$ points, and we perform an incoherent sum of spectra within the mask to acquire a dark-field EELS v-DOS signal. The positions of the masks are shown in Fig. \ref{fig:AlNDOS}.b over the 3 mrad diffraction image, however the aperture-selected v-DOS shown in Fig. \ref{fig:AlNDOS}.c,d are for the parallel illumination case. 

To simulate a comparable signal from MD, a grid of $k$ points are used to calculate  $\vv{\bm{\rho}}(\omega,\vv{\bm{k}})_{coh}$, extending across multiple interferometric Brillouin zones, $\vv{\bm{q}}\bullet\vv{\bm{u}}$ selection rules are applied to reach  $\vv{\bm{\mathcal{D}}}(\omega,\vv{\bm{k}})$, and a similar masking operation is performed. Merely summing the dispersions acquired in the high-symmetry directions is insufficient to replicate the q-EELS signal, as it does not account for all modes across reciprocal space. Similarly, taking the FFT of velocities from MD (related to the v-DOS) in each direction may capture modes at reciprocal points outside of the dark field aperture (mask). Instead, the mask applied to  $\vv{\bm{\rho}}(\omega,\vv{\bm{k}})_{coh}$ instead of  $\vv{\bm{\mathcal{D}}}(\omega,\vv{\bm{k}})$ will provide a polarization-specific ``ground truth'' spectrum for eigenvectors within the selected reciprocal area, with selection rules excluded (shown in Fig. \ref{fig:AlNDOS}.d, dotted)

Starting by comparing the central, right, and upper Brillouin zones, these should have selectivity for longitudinal modes, and vibrations in $x$ and $z$. These are shown in Figure \ref{fig:AlNDOS}.c (black, green, and blue, respectively). Reasonable agreement is seen between the spectrum acquired from q-EELS and $\vv{\bm{\mathcal{D}}}(\omega,\vv{\bm{k}})$ (solid and dashed, respectively).

Total direction selectivity should not be expected however. For a dark field mask centered on a +$x$ $\Gamma$ point (for example), a finite mask radius implies some sensitivity will remain to eigenvector components in the $z$ direction. It may thus be intuitive that higher-order Brillouin zones should yield a better selectivity for a given polarization, as the $z$ eigenvector component is minimized. To explore this, we examine three Brillouin zones in the $\Gamma$-$K$-$M$ direction (ignoring the $\Gamma$-$A$ direction due to the large interferometric Brillouin zone in $z$). These results are shown in Figure \ref{fig:AlNDOS}.d compared against the $x$ direction ``ground-truth'' v-DOS spectrum from $\vv{\bm{\rho}}(\omega,\vv{\bm{k}})_{coh}$ (i.e., with no selection rules included). We actually see a worsening selectivity, likely due to increasing sensitivity to through-plane vibrations at high-$q$ and multiple scattering effects. To support this premise, we point to the diffraction patterns acquired from abtem \cite{abTEM}. abtem captures kinematic and dynamic effects, and faint forbidden reflections can be seen. These effects also become more dominant at higher $q$. A comparison of the abtem diffraction pattern (kinematic and dynamic effects) and that from py4DSTEM \cite{Savitzky2021} (kinematic only) is shown in the Supplemental Material, showing the presence of faint forbidden reflections when dynamic effects are included.

\subsection{Convergent beam density of states}

In theory, direction selectivity should also be reduced with an increasing convergence angle, as additional uncertainty from the electron's incoming momentum means a much larger range of phonon momenta are captured by the dark field aperture or mask. This is shown in Figure \ref{fig:AlNDOS}.e,f for the parallel beam case, and for convergence angles of 1, 3, 5, 15, and 30 mrad. At low convergence angles, direction selectivity is preserved, but it lessens as the convergence angle is increased. For example, there are clear differences in the parallel-beam DOS (black curves) between \ref{fig:AlNDOS}.e and f, particularly in the 15-18 THz range. In contrast, the differences at 30 mrad (blue) are much more subtle. 

Increasing convergence also saw an increasing level of noise in the v-DOS signal when a single probe position is used. Fundamentally, no finite-duration simulation will capture all available states equally, meaning increased noise will be seen for shorter simulations of when fewer atoms are sampled. For a high convergence angle, only a single column of atoms might be sampled, resulting in more noise in the signal. Coherent interference between overlapping Bragg disks (e.g. for adjacent phonon modes with no correlation in phase) will also serve to lessen the total signal. For these reasons, all convergent-beam q-EELS results in this manuscript were prepared by averaging the v-DOS signal from 50 probe positions, with a low-pass filter applied after. While this approach works for bulk materials, samples with localized vibrations can achieve the same effect by sampling over multiple trajectories or a longer duration. Raw v-DOS for a single point and unfiltered v-DOS for each convergence angle are included in the Supplemental Material for reference. While many probe positions or collection over multiple molecular dynamics trajectories is required to mitigate noise issues for convergent-beam q-EELS simulations, this is not necessary in experiment. Realistic acquisition times are on the order of seconds or minutes, as compared to nanosecond-duration molecular dynamics simulations. 

To replicate convergent-beam q-EELS with $\vv{\bm{\mathcal{D}}}(\omega,\vv{\bm{k}})$, and if a spatially-localized vibrational response is of interest, an airy function mask (matching that of the desired probe beam profile) can be used to zero the velocities outside of the probe. When this is done, the resulting energy-resolved diffraction images and v-DOS are in reasonable agreement with q-EELS, however computational limitations related to the collection of trajectories over longer molecular dynamics simulations have limited our ability to explore this in depth.

\subsection{Thickness effects}
Calculations performed via MD and LD may not capture thickness-dependent effects which are known to occur in (scanning) transmission electron microscopy experiments. To evaluate these effects in q-EELS simulations, we prepared a thicker MD simulation (400 $\times$ 8 $\times$ 8 unit cells of silicon, with the multislice simulation propagating the electron wave along the 400 unit cell direction). Convergent STEM probes (30 mrad, 5 mrad, and 3 mrad) are propagated, and the exit wave is recorded as a function of depth. Summing the entire v-DOS spectrum yields a signal intensity as a function of depth, which we have presented in Figure \ref{fig:AlNDOS}.g. To better visualize the depth-distribution of the acquired signal, the derivative of the spectrum intensity is taken (Figure \ref{fig:AlNDOS}.h) to obtain the signal attributed to each slice. Regardless of convergence angle, shallower regions of the sample have a higher contribution to the signal (steeper in Figure \ref{fig:AlNDOS}.g, higher signal per layer in Figure \ref{fig:AlNDOS}.h). We also see the Pendell{\"o}sung effect (periodic oscillations in signal intensity) which comes from dynamic scattering effects. A series of simulations were also run with a finer depth-resolution (0.25 unit cell), shown in the inset of Figure \ref{fig:AlNDOS}.g. Interestingly, a stepped behavior in the 30 mrad case is seen. In the diamond cubic structure in [001], an atomic column has an atom on every other monolayer. If atomic resolution is obtained (as in the 30 mrad case), this feature results in near-zero additional signal for the monolayers without atoms. For the 3 and 5 mrad cases, the loss of atomic resolution yields a blurring of this effect. 

In the q-EELS dispersion in Fig. \ref{fig:dispersions}.a, we also see faint spurious branches which we have not commented on until now. These are the product of increased surface sensitivity, which allows detection of through-plane phonons within the simulation. While these branches are not the focus of this work, we have prepared several additional MD simulations and $\vv{\bm{\rho}}(\omega,\vv{\bm{k}})_{coh}$ calculations (available in the Supplemental Material) to support this claim. A measurement of a single monolayer will be sensitive to waves traveling orthogonal to the measurement plane (similarly to how an antenna picks up signals from many directions). The measurement of the next monolayer will be sensitive to the same wave, however a slight phase shift will be seen. For a measurement with uniform sensitivity through the depth of the system, coherent interference will occur between monolayer-specific signals, and the through-plane wave will not be observed. This same effect also prevents the observation of [110] waves (for example) when analyzing $\vv{\bm{\rho}}(\omega,\vv{\bm{k}})_{coh}$ or SED in the [100] direction. 

Any variation in sensitivity with depth will yield sensitivity to these through-plane modes however. In the Supplemental Material, we show dispersions from thickness-dependent q-EELS. The intensity of these phantom branches lessens as the beam propagates deeper into the sample, supporting the premise that destructive interference suppresses the appearance of these branches. We also show similar results via $\vv{\bm{\rho}}(\omega,\vv{\bm{k}})_{coh}$, using only the upper few monolayers for the calculation, suggesting that these branches are inherent to the system. Ten sets of branches can also be counted, in agreement with the 5 conventional unit cells or 10 primitive unit cell thickness used for the silicon simulations (where $N$ eigenfrequencies exist for a finite-sized system of $N$ repeating units in a given direction). We also reproduce these branches with LD, by calculating across a grid of $k$ points with 10 discrete steps in the $k_z$ direction. 

In practice, a suspended film is less likely to contain through-plane coherent vibrational waves (as opposed to a thin slab with periodic boundary conditions applied). We did however prepare a simulation of suspended silicon and found that internal reflection of through-plane waves may still occur. Surface defects/impurities/distortions are likely to disrupt the reflection of these waves however. These have not been observed in experiment to our knowledge, but assuming adequate measurement sensitivity and the right material system, these should be observable. 

\section{Conclusion}
In this work we have shown the presence of a so-called ``interferometric Brillouin zone'', which is a larger minimum repeating unit in reciprocal space, defined by the interatomic spacing rather than the size of the primitive cell. This phenomena arises from interference of phonons on each atomic basis index, and yields a vibrational non-equivalency between traditional Brillouin zones. The interferometric Brillouin zone has been observed in fast-electron experiments \cite{Li2024}, and the behavior is observed here without considering the incident electron's interaction. 

We also investigated the effects of selection rules \cite{Nicholls2019}, where the fast-electron momentum exchange is directly tied to the direction of atomic displacements (eigenvectors). These result in the disappearance of specific phonon branches within certain Brillouin zones, and this effect can be used to acquire a polarization-selective vibrational density of states (v-DOS) measurement.

Based on our understanding of selection rules, and noting the equivalence of the coherent probability density (a spatio-temporal Fourier transform of atomic displacements, $\vv{\bm{\rho}}(\omega,\vv{\bm{k}})_{coh}$, Eq. \ref{eq:SEDco}) to the single-phonon scattering terms in the dynamic structure factor, we present an analysis of molecular dynamics (MD) and Lattice Dynamics (LD) as tools for reproducing the q-EELS signal to a first approximation. 

Finally, we note several additional effects, including variation in the direction-selective v-DOS based on convergence angle or interferometric Brillouin zone, multiple-scattering effects, a surface-selectivity in q-EELS, and the presence of spurious branches in the q-EELS phonon dispersions due to through-plane modes. 

\section{Acknowledgments}

T.W.P, E.R.H, \& J.A.H. acknowledge the support of the U.S. Department of Energy, Office of Basic Energy Sciences (DOE-BES), Division of Materials Sciences and Engineering under contract ERKCS89.

T.W.P. \& P.E.H. acknowledge instrument support performed as part of user proposal at the Center for Nanophase Materials Sciences (CNMS), which is a US Department of Energy, Office of Science, User Facility. 

Microscopy performed using instrumentation within ORNL’s Materials Characterization Core provided by UT-Battelle, LLC, under Contract No. DE-AC05- 00OR22725 with the DOE and sponsored by the Laboratory Directed Research and Development Program of Oak Ridge National Laboratory, managed by UT-Battelle, LLC, for the U.S. Department of Energy. 

Work at UVA was supported as part of APEX (A Center for Power Electronics Materials and Manufacturing Exploration), an Energy Frontier Research Center funded by the U.S. Department of Energy (DOE), Office of Science, Basic Energy Sciences (BES), under Award \#ERW0345 (computational studies and analysis). 

Work at Vanderbilt was supported by the U.S. Department of Energy, Office of Science User Facility and the U.S. Department of Energy, Office of Basic Energy Sciences (DOE-BES), Division of Materials Sciences and Engineering grant DE-FG02-o0ER464 and by the McMinn Endowment. 

T.W.P. \& P.E.H. acknowledge Research Computing at The University of Virginia for providing computational resources and technical support that have contributed to the results reported within this publication. URL: https://rc.virginia.edu. 

This research used resources of the Compute and Data Environment for Science (CADES) at the Oak Ridge National Laboratory, which is supported by the Office of Science of the U.S. Department of Energy under Contract No. DE-AC05-00OR22725. 

This work was led equally by University of Virginia and Oak Ridge National Laboratory.

Notice: This manuscript has been authored by UT-Battelle, LLC, under contract DE-AC05-00OR22725 with the US Department of Energy (DOE). The US government retains and the publisher, by accepting the article for publication, acknowledges that the US government retains a nonexclusive, paid-up, irrevocable, worldwide license to publish or reproduce the published form of this manuscript, or allow others to do so, for US government purposes. DOE will provide public access to these results of federally sponsored research in accordance with the DOE Public Access Plan (\hyperlink{https://www.energy.gov/doe-public-access-plan}{https://www.energy.gov/doe-public-access-plan}).

H.A.W. \& S.T.P. acknowledge computing resources from the National Energy Research Scientific Computing Center (NERSC), a U.S. Department of Energy Office of Science User Facility located at Lawrence Berkeley National Laboratory, operated under Contract No. DE-AC02-05CH11231.

\bibliographystyle{unsrt}
\bibliography{Refs.bib}

\section{Appendices}
\subsection{Appendix A: Molecular Dynamics Setup}
Molecular dynamics simulations were performed in the Large-scale Atomic/Molecular Massively Parallel Simulator software (LAMMPS \cite{LAMMPS}) on silicon and Aluminum Nitride (AlN). Silicon serves as a model material, as it is well understood, simple to model, with fast and efficient atomic potentials. We use the Stillinger-Weber potential, with a lattice parameter of 5.43729 \AA. For SED and our parallel-beam q-EELS simulations, our MD simulation volume consists of a 50 $\times$ 50 $\times$ 5 conventional unit cell slab (27.19 $\times$ 27.19 $\times$ 2.72 nm) with periodic boundary conditions in all directions. As the primitive cell is half the size of the conventional cell, this yields 100 $k$ points in each in-plane direction within the first Brillouin zone. Given the crystal symmetry of the diamond cubic structure, we evaluated vibrations in the [001] plane (in the [100] direction: $\Gamma-X$ and in [110]: $\Gamma-K$). For our thickness-dependent series of convergent-beam simulations, we prepared a separate simulation of 400 $\times$ 8 $\times$ 8 unit cells (217.5 $\times$ 4.35 $\times$ 4.35 nm). The 8 unit cell simulation width yields very poor reciprocal space resolution, but allowed tracking of the q-EELS v-DOS signal through a large depth. For thickness-dependent q-EELS simulations, the exit wave was exported every 2.5 unit cells (13.6 nm). All silicon simulations used timesteps of 2 fs, equilibrated under NVT for 1 ns (500k steps), and under NVE for 2 ns (1M steps). Following equilibration, positions and velocities were dumped every 20 fs, for an additional 10 ps (500 timesteps). This translates to a maximum measurable frequency of 25 THz ($f_{range}=1/\Delta t$, where an FFT finds both positive and negative frequencies) and a frequency resolution of 0.1 THz ($\Delta \omega=1/duration$). 

For AlN, we used a DFT-trained deepMD potential. This potential has been validated previously 
and it faithfully reproduces the phonon dispersion (with the exception of near-$\Gamma$ optical modes, where the local descriptor fails to capture long-range interactions). For AlN, we evaluated vibrations occurring in the [001] and [010] planes. In the [001] plane, the hexagonal structure was merely used as a demonstration of the interferometric Brillouin zone. The [010] shows a different interferometric Brillouin zone, and also allows for comparison of anisotropic behavior between $\Gamma$-$K$-$M$ and $\Gamma$-$A$ Brillouin zone directions (real-space and reciprocal-space structure shown in the Supplemental MAterial). For the [001] plane simulations, we simulate a structure 50 $\times$ 50 $\times$ 2 unit cells (using lattice constants $a=3.188930$ \AA{} and $c=5.192357$ \AA{}) using a skewed cell (non-orthogonalized). For simulations in the [010] plane, the system is 50 $\times$ 2 $\times$ 31 unit cells (which translates to a roughly-square slab of 15.94 $\times$ 16.10 nm).  


\subsection{Appendix B: Lattice Dynamics Calculations}
We use the same Stillinger-Weber potential and lattice constants used for MD. Supercells with atomic displacements are generated via phonopy \cite{phonopy1,phonopy2}, forces calculated from each via LAMMPS, 2nd order force constants calculated via phonopy, followed by the eigenvectors for arbitrary $k$ points calculated again via phonopy. Code examples are available on the phonopy github, and our code for these calculations is available upon request. 

\subsection{Appendix C: Momentum-Resolved EELS calculations}
Our q-EELS simulations were performed in a manner similar to that of Zeiger \cite{Zeiger2021} and Castellanos-Reyes \cite{CR2025}. In either case, molecular dynamics (MD) simulations are used to acquire time-dependent atomic configurations, which are then used as the input for frozen phonon multislice (electron wave propagation) simulations. While the electron wave simulations are elastic (no energy loss is simulated), the result contains information on the frequency-dependent vibrations from MD and the scattering probabilities of the transmitted electrons \cite{Forbes2010}. 

Two slightly differing methods for these simulations have been developed, the first of which was by Zeiger \textit{et al.} \cite{Zeiger2021}. Given the frequency-resolved nature of these simulations and the use of multislice simulations for the electron wave, they have been referred to as ``Frequency-Resolved Frozen-Phonon Multislice'' or ``FRFPMS''. These simulations originally made use of a custom frequency-specific thermostat to generate atomic configuration snapshots corresponding to a given frequency, however subsequent works simply use a band-pass filter over a single MD simulation. The resulting frequency-filtered atomic configurations are then used for frozen phonon multislice simulations. The q-EELS signal for a given energy bin is then obtained by taking the difference between the incoherent and coherent sum across multiple frozen phonon configurations.

In the work of Castellanos-Reyes \textit{et al.} \cite{CR2025}, multislice simulations are performed over consecutive timesteps, and a Fourier transform over time yields the energy-resolved q-EELS signal. The Fourier transform over the multislice exit wave is referred to as the ``time autocorrelation of the auxiliary wave'' and this approach has thus been referred to as the ``TACAW'' method. TACAW yields similar results to FRFPMS, but it is much more computationally efficient. FRFPMS requires multiple frozen phonon configurations for each energy (where parameters such as convergence-angle and sample thickness affect the number of configurations required). By comparison, the number of time-steps used in TACAW directly translates to the frequency resolution. 

In this work, we use the TACAW method (the post-hoc Fourier transform approach developed by Castellanos-Reyes \textit{et al.} \cite{CR2025}). We also performed simulations using FRFPMS (the frequency-binning method developed by Zeiger \textit{et al.} \cite{Zeiger2021}) to ensure observations were maintained (an example is available in the Supplemental Material). 
None of our findings should be unique to the simulation method (``FRFPMS'' vs. ``TACAW''), or the software packages used (LAMMPS, or abTEM). 

\end{document}


\renewcommand{\thefigure}{S\arabic{figure}}
\renewcommand{\thetable}{S\arabic{table}}
\renewcommand{\theequation}{S\arabic{equation}}

\linespread{1.6}\selectfont{} 
\preprint{JRN/123-ABC}

\title{Phonon selection and interference in momentum-resolved electron energy loss spectroscopy - Supplemental Material}

\author{Thomas W. Pfeifer}
\email{pfeifertw@ornl.gov}
\affiliation{Center for Nanophase Materials Sciences, Oak Ridge National Laboratory, Oak Ridge,  Tennessee 37830, USA}
\affiliation{previously Department of Mechanical and Aerospace Engineering, University of Virginia, Charlottesville, Virginia 22920, USA}

\author{Harrison A. Walker}
\affiliation{Interdisciplinary Materials Science Program, Vanderbilt University, Nashville, Tennessee 37235, USA}

\author{Henry Aller}
\affiliation{Department of Mechanical Engineering, University of Maryland, College Park, Maryland 20742, USA}

\author{Samuel Graham}
\affiliation{Department of Mechanical Engineering, University of Maryland, College Park, Maryland 20742, USA}

\author{Sokrates Pantelides}
\affiliation{Interdisciplinary Materials Science Program, Vanderbilt University, Nashville, Tennessee 37235, USA}
\affiliation{Department of Physics and Astronomy, Vanderbilt University, Nashville, Tennessee 37235, USA}

\author{Jordan A. Hachtel}
\affiliation{Center for Nanophase Materials Sciences, Oak Ridge National Laboratory, Oak Ridge,  Tennessee 37830, USA }

\author{Patrick E. Hopkins}
\email{peh4v@virginia.edu}
\affiliation{Department of Mechanical and Aerospace Engineering, University of Virginia, Charlottesville, Virginia 22920, USA}
\affiliation{Department of Physics, University of Virginia, Charlottesville, Virginia 22920, USA}
\affiliation{Department of Materials Science and Engineering, University of Virginia, Charlottesville, Virginia 22920, USA}

\author{Eric R. Hoglund}
\email{hoglunder@ornl.gov }
\affiliation{Center for Nanophase Materials Sciences, Oak Ridge National Laboratory, Oak Ridge,  Tennessee 37830, USA}

\date{\today}

\maketitle

\newpage

\section{\label{sec:level1} Supplemental Material}

\subsection{Code availability}
Our q-EELS simulations tools are available at \hyperlink{https://github.com/tpchuckles/abEELS}{https://github.com/tpchuckles/abEELS}. This tool accepts an input configuration file (samples available on github as well) which specifies the LAMMPS output files, atom types, system size and timestep information. Optional system trimming/tiling/rotation parameters are also included. Both FRFPMS and TACAW can be run using the same code. Generation of frequency bins (for FRFPMS) is performed, and the multislice calculation is done via abtem \cite{abTEM}. Post-processing tools are included: v-DOS, phonon dispersions via traces along reciprocal paths, or energy-resolved diffraction images. 

Our coherent probability density code is available at \hyperlink{https://github.com/tpchuckles/pySED}{https://github.com/tpchuckles/pySED}. Common functions are available, which can be imported into any python script. Example scripts are also included in the ``examples'' folder, showing generation of dispersions in various materials and reciprocal directions. 

\subsection{Comparing FRFPMS and TACAW}
\begin{figure}
    \centering
    \includegraphics[width=0.95\linewidth]{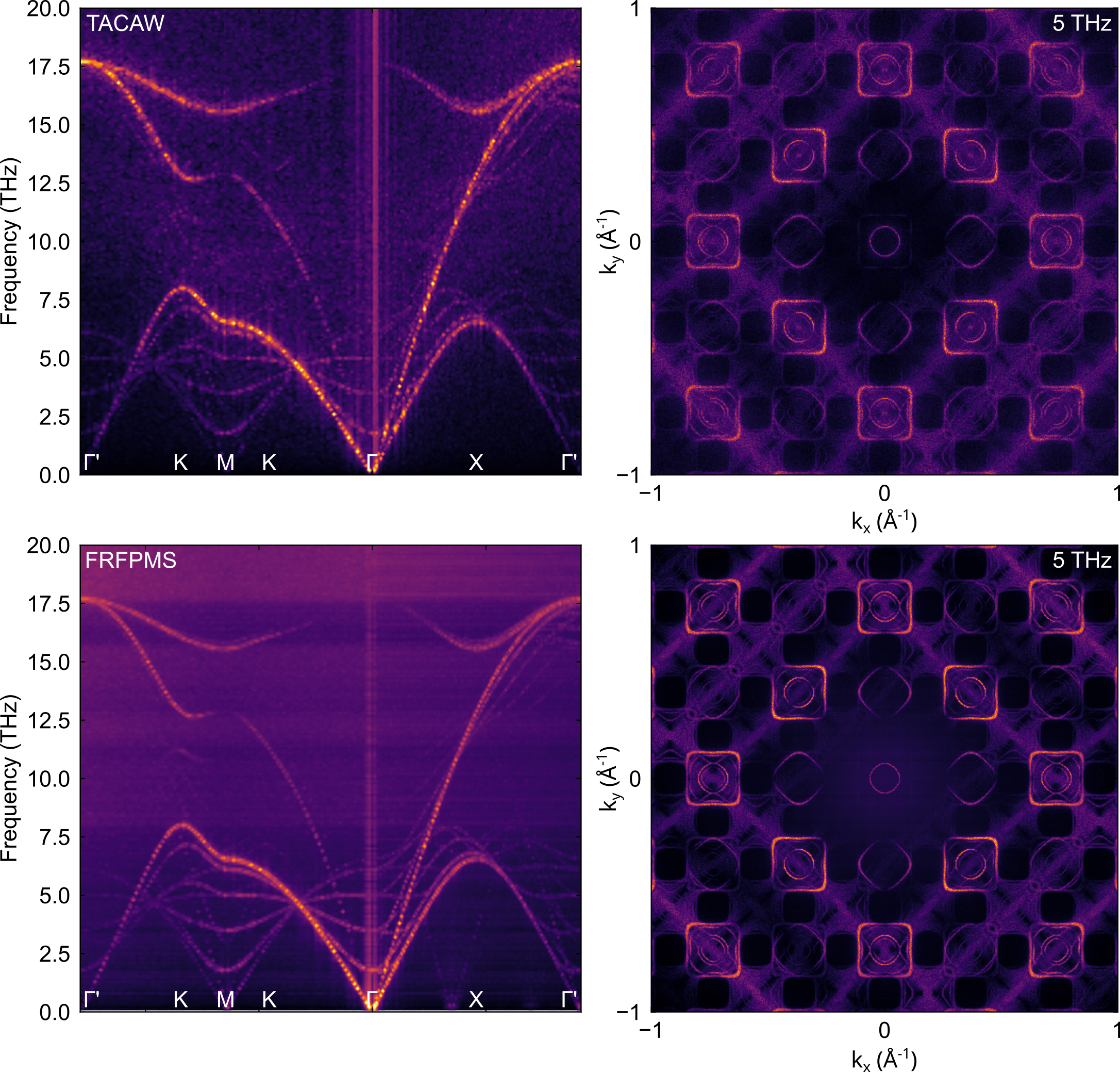}
    \caption{The two methods for q-EELS simulations are compared: FRFPMS developed by Zeiger \textit{et al.} and TACAW developed by Castellanos-Reyes \textit{et al.} \cite{CR2025}. All features in the data appear to be consistent regardless of which method is used.}
    \label{fig:PZJACR}
\end{figure}

We (and previous authors) have found that the two q-EELS simulations methods: FRFPMS developed by Zeiger \textit{et al.} and TACAW developed by Castellanos-Reyes \textit{et al.} \cite{CR2025}, yield similar results. We thus use the more computationally-efficient TACAW method for the majority of the q-EELS simulation in this work, however we provide a brief comparison between FRFPMS and TACAW here. We have included the parallel beam silicon dispersion and an energy slice for each in Figure \ref{fig:PZJACR}. Our TACAW calculation is performed over 500 consecutive timesteps. Our FRFPMS calculation was performed using 200 bins (every 0.1 THz, spanning a range of 20 THz), with a gaussian frequency filter width of 0.04 THz over 2000 timesteps. 20 randomized frozen phonon configurations were used for each frequency bin. The apparently-higher resolution for FRFPMS is merely a product of the exceptionally-small frequency bin width, and results may vary. 

\subsection{A mathematical comparison between conventions for the phonon eigenvector}

In the main manuscript, we briefly mention the two conventions for the phonon eigenvector. The eigenvector itself is used to determine the wavelike vibrations of atoms, according to either of the following expressions:

\begin{equation}
	\vv{\bm{u}}_{n,j,l,m}(t)=
	A_{j,l,m} \; \vv{\bm{\varepsilon}}_{j,l,m} \; e^{ i \: ( \vv{\bm{k}}_l \bullet \vv{\bm{r}}_{n,j} -  \omega_m \, t)} \;\;\;\; or \;\;\;\; \vv{\bm{u}}_{n,j,l,m}(t)=
	A_{j,l,m} \; \vv{\bm{\varepsilon}}_{j,l,m} \; e^{ i \: ( \vv{\bm{k}}_l \bullet \vv{\bm{r}}_{n} -  \omega_m \, t)}
\end{equation}

where indices $n$ and $j$ denote an individual atom (residing in unit cell $n$ at basis index $j$), and indices $l$ and $m$ denote an individual mode (i.e. the vibration at wavevector $\vv{\bm{k}}_l$ and frequency $\omega_m$). 

The key difference is in the use of the atomic position vector $\vv{\bm{r}}_{n,j}$, or the use of the vector to the unit cell origin $\vv{\bm{r}}_{n}$. These differing conventions are discussed most clearly in Srivastava \cite{Srivastava2022} Equations 2.48-2.60. In traditional lattice dynamics, the D-Type dynamical matrix is used to calculate eigenvectors following a convention using unit cell positions $\vv{\bm{r}}_{n,j=0}$, whereas the C-Type dynamical matrix uses $\vv{\bm{r}}_{n,j}$. The resulting eigenvectors will have a phase shift according to $e^{i \, \vv{\bm{k}}\bullet\vv{\bm{s}}_j}$, if the atom's relative position within the unit cell $\vv{\bm{s}}_j$ is defined by $\vv{\bm{r}}_{n,j}=\vv{\bm{r}}_{n,j=0}+\vv{\bm{s}}_{j}$. Using a different dynamical matrix will result in the calculation of different values of $\vv{\bm{\varepsilon}}_{j,l,m}$, compatible with the differing forms of expressions above. 

This is particularly relevant when considering the summation of eigenvectors for the coherent probability density, and the inclusion or exclusion of this additional phase shift term also depends on our understanding of the meaning of the summed eigenvector.

While the per-basis eigenvector $\vv{\bm{\varepsilon}}_{j,l,m}$ defines the wave which describes the motion of only the atoms at $j$ basis index, the summation of eigenvectors $\vv{\bm{\varepsilon}}_{s,l,m}$ can be understood as defining the \textit{singular} wave which approximately describes the atomic motion across \textit{all} atoms. 

Two conventions may exist for this wave as well, and the choice of convention for $\vv{\bm{\varepsilon}}_{j}$ and separately for $\vv{\bm{\varepsilon}}_{s}$ will determine the necessity of the additional phase term:

\begin{equation}
	\vv{\bm{u}}_{n,j,l,m}(t)=
	A_{s,l,m} \; \vv{\bm{\varepsilon}}_{s,l,m} \; e^{ i \: ( \vv{\bm{k}}_l \bullet \vv{\bm{r}}_{n,j} -  \omega_m \, t)} \;\;\;\; or \;\;\;\; \vv{\bm{u}}_{n,j,l,m}(t)=
	A_{s,l,m} \; \vv{\bm{\varepsilon}}_{s,l,m} \; e^{ i \: ( \vv{\bm{k}}_l \bullet \vv{\bm{r}}_{n} -  \omega_m \, t)}
\end{equation}

In the following examples, several simplifications will be made for brevity. Time will be neglected (or the wave will be analyzed at $t=0$). Displacements will be considered in 1D (all vector notation will be removed, the displacements will represent a single phonon polarization in a single direction). Amplitude terms will be neglected. 

We will be considering a single wavelength, $k_l$, for atoms indexed by their unit cell $n$, and basis index within that unit cell $j$. 

\subsubsection{$r_{nj}$ convention throughout}

Per our first convention, for a given basis index $j$, the real-space displacements are found from $\varepsilon_j$ via the real-space \textit{atomic} positions $r_{n,j}$, as opposed to the \textit{unit cell} positions $r_n$:

\begin{equation}
	u_{n,j,l}(t) = \varepsilon_{j,l} \; e^{i \, k_l \, r_{n,j} }
	\label{eq:unj}
\end{equation}

where this expression describes a different wave for $j=1,2...$, for as many atoms are in the basis. 

We will also define a similar form for the singular wave attempting to describe the motion of all atoms: 

\begin{equation}
	u_{n,j,l}(t) = \varepsilon_{s,l} \; e^{i \, k_l \, r_{n,j} }
	\label{eq:unj}
\end{equation}

In both cases, the Fourier transform used to determine $\varepsilon$ will use the atomic position (not unit cell position):

\begin{equation}
	\varepsilon_{j,l} = \frac{1}{N} \sum_{n}^{N} u_{n,j}(t) \; e^{-i \, k_l \, r_{n,j} }
\end{equation}

\begin{equation}
	\varepsilon_{s,l} = \frac{1}{N\,B}\sum_{n,j}^{N,B} u_{n,j}(t) \; e^{-i \, k_l \, r_{n,j} }
\end{equation}

The same form for the Fourier transform thus applies for both $\varepsilon_{j,l}$ if we use only the atoms at basis index $j$, or for $\varepsilon_{s,l}$ if we sum over all atoms in the system. 

Considering the case of a 2-atom basis for simplicity, $j=1$ and $j=2$, our above definitions should make it clear that $\varepsilon_{s,l}=\frac{1}{B} \sum_j^B \varepsilon_{j,l}$: 

\begin{equation}
	\varepsilon_{s,l} = 
	\frac{1}{2}(\varepsilon_{1,l} + \varepsilon_{2,l}) =
	\frac{1}{N\,B}\sum_{n,j}^{N,B} u_{n,j}(t) \; e^{-i \, k_l \, r_{n,j} } =
	\frac{1}{2} \left(
	\frac{1}{N} \sum_{n}^{N} u_{n,1}(t) \; e^{-i \, k_l \, r_{n,1} } +
	\frac{1}{N} \sum_{n}^{N} u_{n,2}(t) \; e^{-i \, k_l \, r_{n,2} } \right)
\end{equation}

In an effort to relate this to the $r_n$ convention, it may be intuitive to separate $r_{n,j}$ into a unit cell position $r_n$, and the atom's relative position from the unit cell origin: $s_j$, i.e.: 

\begin{equation}
	r_{n,j}=r_n+s_j
\end{equation}

In this case, the above equations can be rewritten. Noting that when $j=1$, $s_j$ is likely zero, we find:

\begin{equation}
	u_{n,1,l}(t) = \varepsilon_{1,l} \; e^{i \, k_l \, r_{n} } \;\;\;\; and \;\;\;\; u_{n,2,l}(t) = \varepsilon_{2,l} \; e^{i \, k_l \, r_{n} } \; e^{i \, k_l \, s_{2} }
	\label{eq:unj}
\end{equation}

There is an additional $e^{i \, k_l \, s_{2} }$ phase shift term which appears when $j\neq1$. Plugging the second expression back into our defined Fourier transform however:

\begin{equation}
	\varepsilon_{2,l} = \frac{1}{N} \sum_{n}^{N} 
	\varepsilon_{2,l} \; e^{i \, k_l \, r_{n} } \; e^{i \, k_l \, s_{2} }
	\; e^{-i \, k_l \, r_{n,j} }
\end{equation}

we must split out the Fourier transform's $r_{n,j}$ term as well:

\begin{equation}
	\varepsilon_{2,l} = \frac{1}{N} \sum_{n}^{N} 
	\varepsilon_{2,l} \; e^{i \, k_l \, r_{n} } \; e^{i \, k_l \, s_{2} }
	\; e^{-i \, k_l \, r_{n} } \; e^{-i \, k_l \, s_{2} }
\end{equation}

and we arrive at the same result. This means the summation of $\varepsilon_{s,l} = 
\frac{1}{2}(\varepsilon_{1,l} + \varepsilon_{2,l})$ shown previously is maintained.

\subsubsection{$r_{n}$ convention for $\varepsilon_j$, but an $r_{n,j}$ convention for $\varepsilon_s$}

We now define the individual waves according to:

\begin{equation}
	u_{n,j,l}(t) = \varepsilon_{j,l} \; e^{i \, k_l \, r_{n} }
	\label{eq:unj}
\end{equation}

as this convention is more common in the literature. We will maintain our previous definition for the "singular wave describing the motion of all atoms" however: 

\begin{equation}
	u_{n,j,l}(t) = \varepsilon_{s,l} \; e^{i \, k_l \, r_{n,j} }
	\label{eq:unj}
\end{equation}

The fourier transforms for $\varepsilon$ will thus differ depending on whether we are looking for $\varepsilon_j$ or $\varepsilon_s$:

\begin{equation}
	\varepsilon_{j,l} = \frac{1}{N} \sum_{n}^{N} u_{n,j}(t) \; e^{-i \, k_l \, r_{n} }
\end{equation}

\begin{equation}
	\varepsilon_{s,l} = \frac{1}{N\,B}\sum_{n,j}^{N,B} u_{n,j}(t) \; e^{-i \, k_l \, r_{n,j} }
\end{equation}

It is now required to pull out the phase factor from $s_2$, in order to accommodate the differing conventions of $r_n$ and $r_{n,j}$:

\begin{equation}
	u_{n,2,l}(t) = \varepsilon_{2,l} \; e^{i \, k_l \, r_{n} } = \varepsilon_{2,l} \; e^{i \, k_l \, r_{n,j} } 
	\;\;\;\; and \;\;\;\;
	u_{n,2,l}(t) = \varepsilon_{2,l} \; e^{i \, k_l \, r_{n} } = \varepsilon_{2,l} \; e^{i \, k_l \, r_{n,j} } \; e^{-i \, k_l \, s_{2} }
\end{equation}

and $\varepsilon_{s,l}$ does not come from the mere summation of $\varepsilon_{1,l}$ and $\varepsilon_{2,l}$:

\begin{equation}
	\varepsilon_{s,l} = 
	\frac{1}{2}(\varepsilon_{1,l} + \varepsilon_{2,l} \; e^{-i \, k_l \, s_{2} }) =
	\frac{1}{N\,B}\sum_{n,j}^{N,B} u_{n,j}(t) \; e^{-i \, k_l \, r_{n,j} } =
	\frac{1}{2}
	\frac{1}{N} \sum_{n}^{N} \left(
	\varepsilon_{1,l} e^{i \, k_l \, r_{n,j} }
	+ 
	\varepsilon_{2,l}  e^{i \, k_l \, r_{n,j} } e^{-i \, k_l \, s_{2} } \right) e^{-i \, k_l \, r_{n,j} }
\end{equation}

\subsubsection{$r_{n}$ convention for both $\varepsilon_j$ and $\varepsilon_s$}

Finally, we move to the $r_n$ convention for both:

\begin{equation}
	u_{n,j,l}(t) = \varepsilon_{j,l} \; e^{i \, k_l \, r_{n} }
	\label{eq:unj}
\end{equation}

and 

\begin{equation}
	u_{n,j,l}(t) = \varepsilon_{s,l} \; e^{i \, k_l \, r_{n} }
	\label{eq:unj}
\end{equation}

with a consistent (although atypical) definition for the fourier transform, using the unit cell position rather than atomic position:

\begin{equation}
	\varepsilon_{j,l} = \frac{1}{N} \sum_{n}^{N} u_{n,j}(t) \; e^{-i \, k_l \, r_{n} }
\end{equation}

\begin{equation}
	\varepsilon_{s,l} = \frac{1}{N\,B}\sum_{n,j}^{N,B} u_{n,j}(t) \; e^{-i \, k_l \, r_{n} }
\end{equation}

we again find the summation without additional phase shifts required:

\begin{figure}
	\centering
	\includegraphics[width=0.95\linewidth]{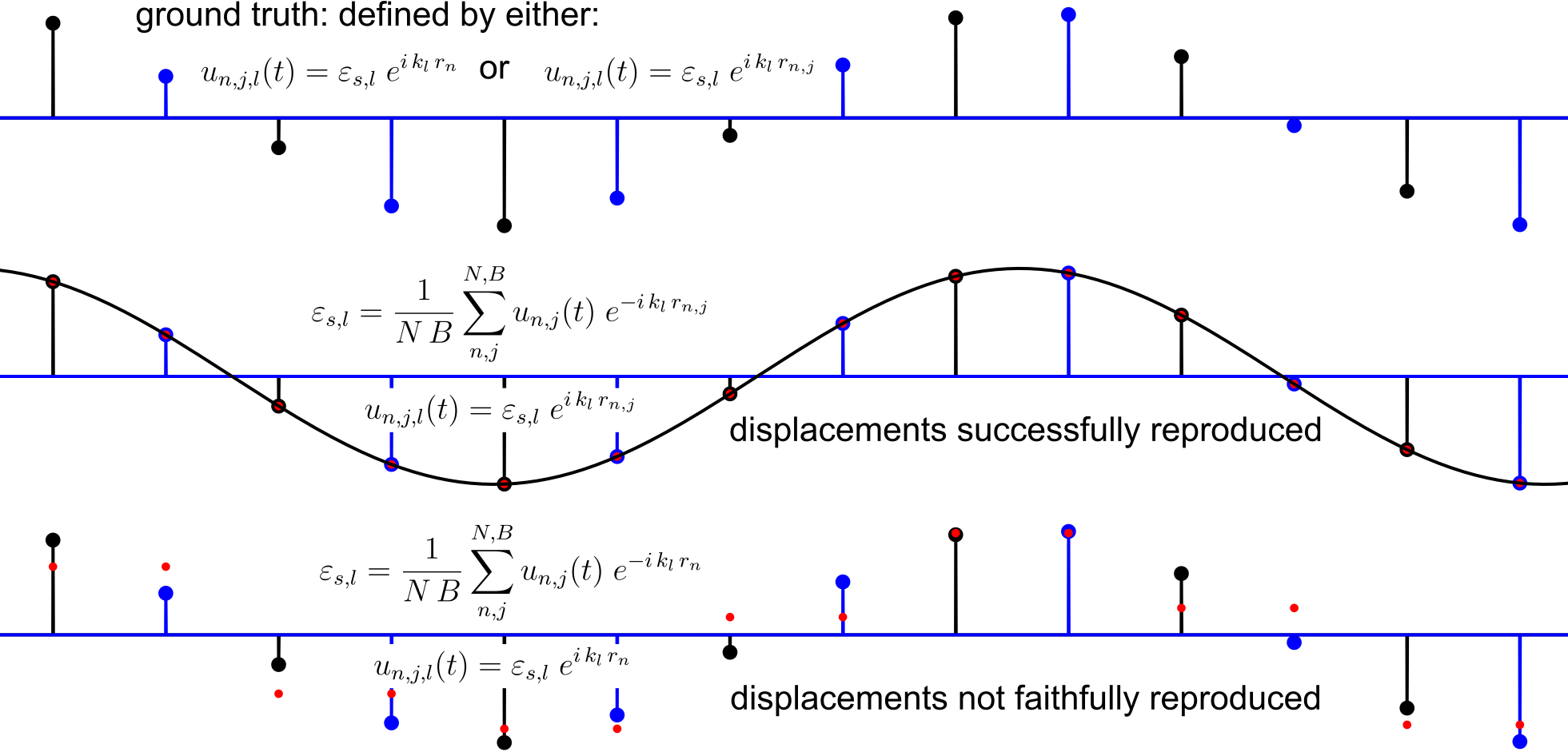}
	\caption{When the $r_{n,j}$ convention is used for both $\varepsilon_{j}$ and  $\varepsilon_{s}$, no phase correction is required and $\varepsilon_{s}=\frac{1}{B} \sum_j^B \varepsilon_{j}$. The atomic displacements over \textit{all} atoms are successfully reproduced by the \textit{singular} wave described by $\varepsilon_{s}$ (middle, solid sinusoid and red dots). When the $r_{n}$ convention is used for $\varepsilon_{j}$ but the $r_{n,j}$ convention is used for $\varepsilon_{s}$, a phase correction term is required: $\varepsilon_{s}=\frac{1}{B} \sum_j^B \varepsilon_{j} \; e^{-i\,k_l,s_j}$, however the atomic displacements over \textit{all} atoms are still successfully reproduced via $\varepsilon_{s}$. If the  $r_{n}$ convention is used by $\varepsilon_{s}$ (bottom), the atoms within the unit cell are treated as existing at the same position, and an effective unit-cell mean displacements are computed, which fail to fully reproduce the displacements of the original system.}
	\label{fig:rnj}
\end{figure}

\begin{equation}
	\varepsilon_{s,l} = 
	\frac{1}{2}(\varepsilon_{1,l} + \varepsilon_{2,l}) =
	\frac{1}{N\,B}\sum_{n,j}^{N,B} u_{n,j}(t) \; e^{-i \, k_l \, r_{n} } =
	\frac{1}{2} \left(
	\frac{1}{N} \sum_{n}^{N} u_{n,1}(t) \; e^{-i \, k_l \, r_{n} } +
	\frac{1}{N} \sum_{n}^{N} u_{n,2}(t) \; e^{-i \, k_l \, r_{n} } \right)
\end{equation}

The meaning of $\varepsilon_{s,l}$ and its accuracy in this scenario is worth considering however. A singular $\varepsilon_{s,l}$ is used, and the \textit{same} spatial coordinate is used for determining the displacements for atoms at $j=1$ and $j=2$. 

Visualizing the displacements resulting from the $r_{n}$ convention used for $\varepsilon_{s,l}$ should make this issue apparent, shown in Fig. \ref{fig:rnj}. 

A similar exercise can be performed for the remaining combination of conventions: $r_{n,j}$ for $\varepsilon_j$ but $r_{n}$ for $\varepsilon_s$, however we highlight the loss of physical meaning in either case, when $\varepsilon_s$ uses the $r_{n}$ convention. The concepts of coherent summation may be maintained, but the inability to tie $\varepsilon_s$ to physical real-world real-space displacements is lost. 

While the above figure shows an acoustic mode (where $\varepsilon_s=\varepsilon_{j=1}=\varepsilon_{j=2}$), the math above holds for optic modes. Where phase cancellation occurs at a given $k_l$, $\varepsilon_s$ may be reduced in amplitude, and will no longer exactly describe the displacements of all atoms. Instead, the full Fourier transform across all $k$ will a $\varepsilon_s$ at a different wavevector which describes the motion. Thus the concept of Brillouin zone unfolding (or the interferometric Brillouin zone) is captured. 

\subsection{A mathematical comparison between incoherent probability density and Spectral Energy Density}

Recall we arrived at the following the incoherent probability density:

\begin{equation} \label{eq:SEDinco} 
	\vv{\bm{\rho}}_{inco}(\omega,\vv{\bm{k}})
	=
	\frac{1}{N \, \tau_f }
	\sum_{j}^{B}
	\bigg|
	\int_{0}^{\tau_{f}} 
	\sum_{n}^{N_u}
	\vv{\bm{u}}_{n,j}(t) \;
	e^{ - i \, ( \vv{\bm{k}} \bullet \vv{\bm{r}}_{n,j} - \omega \, t ) } dt
	\bigg|^{2}
\end{equation}

where the atom is referenced by its unit cell ($n$) and basis index within the unit cell ($j$), with the displacement vector of atom $n$,$j$ at a given timestep given by $\vv{\bm{u}}_{,n,b}(t)$. This came from a simple discrete Fourier transform in space: $X_k=\frac{1}{N}\sum x_n \; e^{-i \, k \, n}$, and continuous Fourier transform in time: $\mathscr{F}(\omega)=\frac{1}{\tau_f}\int_{0}^{\tau_f} f(t) \; e^{i \, \omega \, t} dt$:. 

We present the expression for Spectral Energy Density, as presented by Thomas \textit{et al.} below, with slight changes to notation for consistency:

\begin{equation} \label{eq:SEDinco} 
	\Phi_{\beta}(\omega,\vv{\bm{k}})
	= \zeta
	\sum_{\alpha}
	\sum_{j}^{B} m_j
	\bigg|
	\int_{0}^{\tau_{f}} 
	\sum_{n}^{N_u}
	v_{\alpha,n,j}(t) \;
	e^{ i \, k_{\beta} \, r_{\beta,n,j=0} - i \, \omega \, t } dt
	\bigg|^{2}
\end{equation}

where we have put all scaling terms into $\zeta=\frac{1}{4 \pi \tau_{f} N_{T}}$. $m_j$ is the mass of the atom at basis index $j$, and the velocity $v$ is used instead of displacements $u$. $\alpha$ denotes the direction of velocity, while $\beta$ represents the wave direction in real/reciprocal space. Directions are thus analyzed individually, and magnitude scalers are used instead of vectors. Multiple phonon polarizations (i.e. longitudinal/transverse branches) are summed after the magnitude is taken (i.e., incoherently), meaning degenerate branches should not serve to subtract (via destructive interference) from the total energy. 

$r_{\beta,n,j=0}$ is the position projected along the direction of $\vv{\bm{k}}$ or along $k_\beta$, which is nearly identical to the dot product $\vv{\bm{k}}\bullet\vv{\bm{r}}$ used previously, however SED uses the unit cell position (indexing $n,j=0$) instead of atomic position (indexing $n,j$). This is due to the use of differing conventions, discussed most clearly in Srivastava \cite{Srivastava2022} Equations 2.48-2.60. In traditional lattice dynamics, the D-Type dynamical matrix is used to calculate eigenvectors following a convention using unit cell positions $\vv{\bm{r}}_{n,j=0}$, whereas the C-Type dynamical matrix uses $\vv{\bm{r}}_{n,j}$. The resulting eigenvectors will have a phase shift according to $e^{i \, \vv{\bm{k}}\bullet\vv{\bm{s}}_j}$, if the atom's relative position within the unit cell $\vv{\bm{s}}_j$ is defined by $\vv{\bm{r}}_{n,j}=\vv{\bm{r}}_{n,j=0}+\vv{\bm{s}}_{j}$. Comparing SED vs the incoherent probability density, the choice of convention does not matter, as the magnitude of taken prior to summation across $j$, meaning phase information is lost. 


Finally, the use of velocity as opposed to displacements is understood as SED representing an energy density, i.e., kinetic energy classically is $E_k=\frac{1}{2}\, m \, v^2$. If we assume periodic oscillatory displacements following $u(t)=e^{-i \, \omega \, t}$ then the two are related, as $v(t)=\frac{du(t)}{dt}=-i \, \omega \, e^{-i \, \omega \, t}$, i.e., there is simply a factor of $\omega$ scaling. 

It should thus be apparent that our vector representation for the incoherent probability density $\vv{\bm{\rho}}_{inco}(\omega,\vv{\bm{k}})$ is functionally identical, aside from the vectorized treatment and scaling terms, to SED's $\Phi_{\beta}(\omega,k)$, despite SED being derived from careful tracking of kinetic energy (classically $E_k=\frac{1}{2} \, m \, v^2$).

\subsection{Additional examples of the interferometric Brillouin zone}

\begin{figure}
    \centering
    \includegraphics[width=0.95\linewidth]{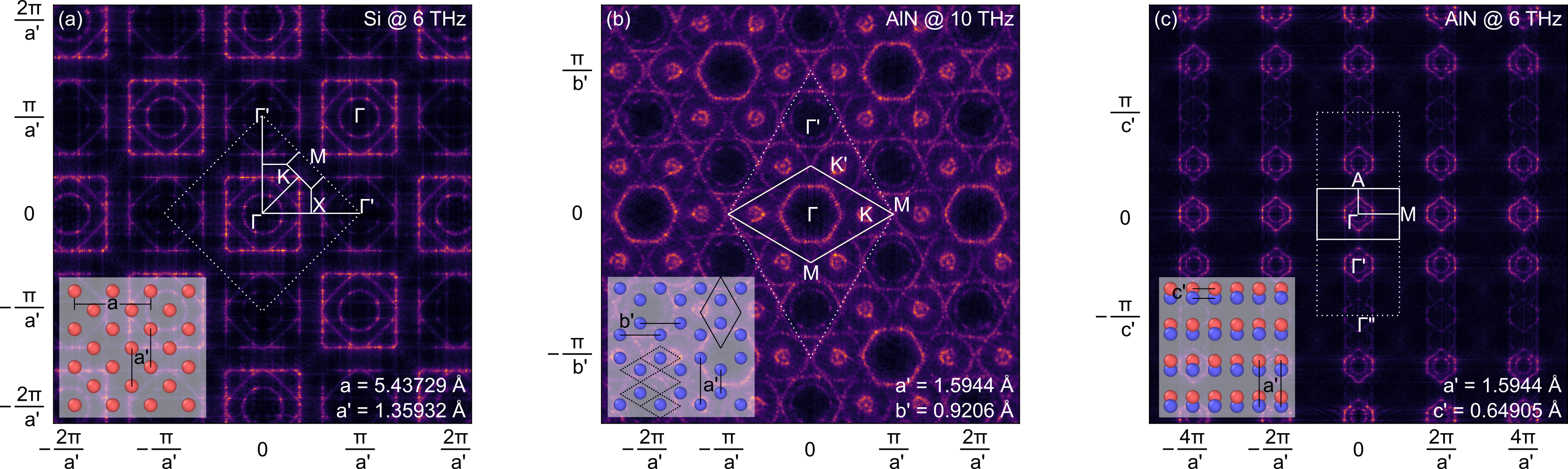}
    \caption{Iso-energy surfaces are shown for (a) silicon in the [001] plane (showing the $\Gamma$-$X$ and $\Gamma$-$K$-$M$ directions) (b) AlN in [001] ($\Gamma$-$K$-$M$ and $\Gamma$-$M$) and (c) AlN in [010] ($\Gamma$-$K$-$M$ and $\Gamma$-$A$). The real-space atomic configurations are shown inset, with conventional Brillouin zones shown in solid lines, and the interferometric Brillouin zones shown dotted. Missing atoms do not affect the interferometric Brillouin zone, as only the interatomic spacing determines the minimum sampling of the vibrational wave.}
    \label{fig:moreIBZs}
\end{figure}

In the main manuscript, we showed the interferometric Brillouin zone for silicon in the [001] plane, however these concepts are not limited to silicon. We thus show the interferometric Brillouin zone for AlN (Wurtzite) in [001] and [010] planes in Fig. \ref{fig:moreIBZs}. For silicon, we saw the interferometric Brillouin zone extend out to $\Gamma'$, however for AlN [001] the interferometric Brillouin zone contains $\Gamma'$ and extends to $K''$. The interferometric Brillouin zone corresponds to a rhombus drawn around a single atom in real space, and some of these cells may be missing atoms, as was the case with silicon. In the [010] direction of AlN, the interferometric Brillouin zone is approximately 4$\times$ as large, due to the closer apparent interatomic spacing in the $c$ direction. 

\subsection{A mathematical comparison between coherent probability density and the Dynamic Structure Factor}

We begin with the definition of the Dynamic Structure Factor from Sturm \cite{Sturm1993} Eq. 22:
\begin{equation}
	\frac{1}{\hbar} S(\bm{k},\omega) = \frac{1}{2\pi\hbar N}
	\int dt \; e^{i \, \omega \, t} 
	\langle \hat{n}(\bm{k},t) \hat{n}(-\bm{k},t=0)\rangle_T
	\label{eq:Sturm22}
\end{equation}
where $\hat{n}(\bm{k},t)$ is defined by Sturm \cite{Sturm1993} Eq 19:
\begin{equation}
	\hat{n}(\bm{k},t)=\sum_j e^{-i \, \bm{k} \bullet r_j(t)}
	\label{eq:Sturm19}
\end{equation}
where $r_j(t)$ is the time-dependent position of atom $j$. 


We will seek a solution in the form of displacements, so we expand $r_j(t)=R_j+u_j(t)$:

\begin{equation}
	\hat{n}(\bm{k},t)=\sum_j e^{-i \, \bm{k} \bullet (R_j+u_j(t))}=\sum_j e^{-i \, \bm{k} \bullet R_j} \; e^{-i \, \bm{k} \bullet u_j(t))}
	\label{eq:Sturm19}
\end{equation}

Note that if $\bm{k} \bullet u_j(t)$ is small, we can take the first terms from a Taylor Expansion:

\begin{equation}
	f(x)=\sum^\infty_{m=0}\frac{1}{m!}\left( f^{(m)}(a) \; (x-a)^m\right)
\end{equation}

where $f^{(m)}(a)$ denotes the $m^{th}$ derivative of the function $f(x)$, evaluated at $x=a$. $e^{-i \, \bm{k} \bullet u_j(t))}$ thus expands to:

\begin{equation}
	e^{-i \, \bm{k} \bullet u_j(t))} \approx
	1 + (-i \, \bm{k} \bullet u_j(t)) + ...
\end{equation}

We note the first term, when plugged in alone, will result in an expression for $S(\bm{k},\omega)$ with no time-dependent terms. We thus focus on the second term only. Ashcroft \& Mermin make a similar expansion in their Appendix N while discussing neutron scattering \cite{ashcroft1976}, and they identify the terms in the taylor expansion as representing the 0, 1,...n phonon scattering terms. We are thus focusing on the single-phonon scattering contribution to the dynamic structure factor. 

We therefore plug:
\begin{equation}
	\hat{n}(\bm{k},t) \approx \sum_j -i \, \bm{k} \bullet u_j(t) \; e^{-i \, \bm{k} \bullet R_j}
	\label{eq:Sturm19}
\end{equation}
back into $S(\bm{k},\omega)$:
\begin{equation}
	\frac{1}{\hbar} S(\bm{k},\omega) \approx \frac{1}{2\pi\hbar N}
	\int_{-\infty}^{\infty}
	\sum_j e^{-i \, 2 \, \bm{k} \bullet R_j} \; e^{i \, \omega \, t} 
	\langle 
	\left(
	-i \, \bm{k} \bullet u_j(t)
	\right)
	\left(
	-i \, \bm{k} \bullet u_j(t=0)
	\right)
	\rangle_T dt
	\label{eq:STATIC}
\end{equation}

and noting the Fourier convolution theorum:
\begin{equation}
	\langle f(t) \; g(t) \rangle = \mathscr{F}^{-1} \left[ \mathscr{F}(f(t)) \; \mathscr{F}(g(t)) \right]
	\;\;\;\;\;or\;\;\;\;\;
	\mathscr{F} \langle f(t) \; g(t) \rangle = f(\omega) \; g(\omega)
	\label{eq:FC}
\end{equation}
then the correlation term is replaced with:
\begin{equation}
	\langle 
	\left(
	-i \, \bm{k} \bullet u_j(t)
	\right)
	\left(
	-i \, \bm{k} \bullet u_j(t=0)
	\right)
	\rangle_T = 
	\mathscr{F}^{-1}\left[
	\mathscr{F}[-i \, \bm{k} \bullet u_j(t)]^2
	\right]
\end{equation}

We also note the definition of the Fourier transform, using the $+i \, \omega \, t$ sign convention as used by Sturm \cite{Sturm1993}. 


\begin{equation}
	\mathscr{F}[f(t)] = \int_{-\infty}^{\infty} f(t) \; e^{i \, \omega \, t} dt
\end{equation}

we return to $S(\bm{k},\omega)$:






\begin{equation}
	\frac{1}{\hbar} S(\bm{k},\omega) \approx \frac{1}{2\pi\hbar N}
	\sum_j e^{-i \, 2 \, \bm{k} \bullet R_j} \;
	\mathscr{F}[-i \, \bm{k} \bullet u_j(t)]^2
	\label{eq:STATIC}
\end{equation}

or, rearranging terms, we arrive at something close to our expression for the coherent probability density:

\begin{equation}
	\frac{1}{\hbar} S(\bm{k},\omega) \approx \frac{1}{2\pi\hbar N} \left[
	\int_{\infty}^{\infty}
	\sum_j
	-i \, \vv{\bm{k}} \bullet \vv{\bm{u}}_j(t)
	e^{-i \, \vv{\bm{k}} \bullet \vv{\bm{R}}_j}
	\; e^{i \, \omega \, t} dt
	\right]^2
	\label{eq:STATIC}
\end{equation}

where the only difference between $\vv{\bm{\rho}}_{coh}(\omega,\vv{\bm{k}})$ (copied from the main text for convenience) is the $\vv{\bm{k}}\bullet\vv{\bm{u}}$ term: 

\begin{equation} \label{eq:SEDco} 
	\vv{\bm{\rho}}_{coh}(\omega,\vv{\bm{k}})
	=
	\frac{1}{N \, \tau_f}
	\bigg|
	\int_{0}^{\tau_{f}} 
	\sum_{j}^{B}
	\sum_{n}^{N_u}
	\vv{\bm{u}}_{n,j}(t) \;
	e^{ - i \, ( \vv{\bm{k}} \bullet \vv{\bm{r}}_{n,j} - \omega \, t ) } dt
	\bigg|^{2}
\end{equation}

Finally, noting that the internal terms from above contain embedded eigenvector information:

\begin{equation}
\vv{\bm{u}}_j(\vv{\bm{k}},\omega) = A_j(\vv{\bm{k}},\omega) \, \vv{\bm{\varepsilon}}_j(\vv{\bm{k}},\omega) =
\frac{1}{N \, \tau_f}
\sum_{n}^{N} \int_{0}^{\tau_f} \vv{\bm{u}}_{n,j}(t) \;
e^{- i \, ( \vv{\bm{k}} \bullet \vv{\bm{r}}_{n,j} - \omega \, t)} dt 
\end{equation}

We see the additional $\vv{\bm{k}}$ dot product and dynamic structure factor can be written in terms of the eigenvectors:
\begin{equation}
S(\bm{k},\omega) \propto \vv{\bm{k}} \bullet \vv{\bm{\varepsilon}}
\end{equation}
which appears in numerous other works \cite{Nicholls2019,Sinha2001,ashcroft1976,Burkel2001}.

\subsection{Additional q-EELS slices for silicon} 

\begin{figure*}[!tp]
	\caption{Energy-resolved diffraction images are shown for silicon for 0.5 THz increments, for a 0-20 THz range}
	\centering
	\includegraphics[width=0.95\linewidth]{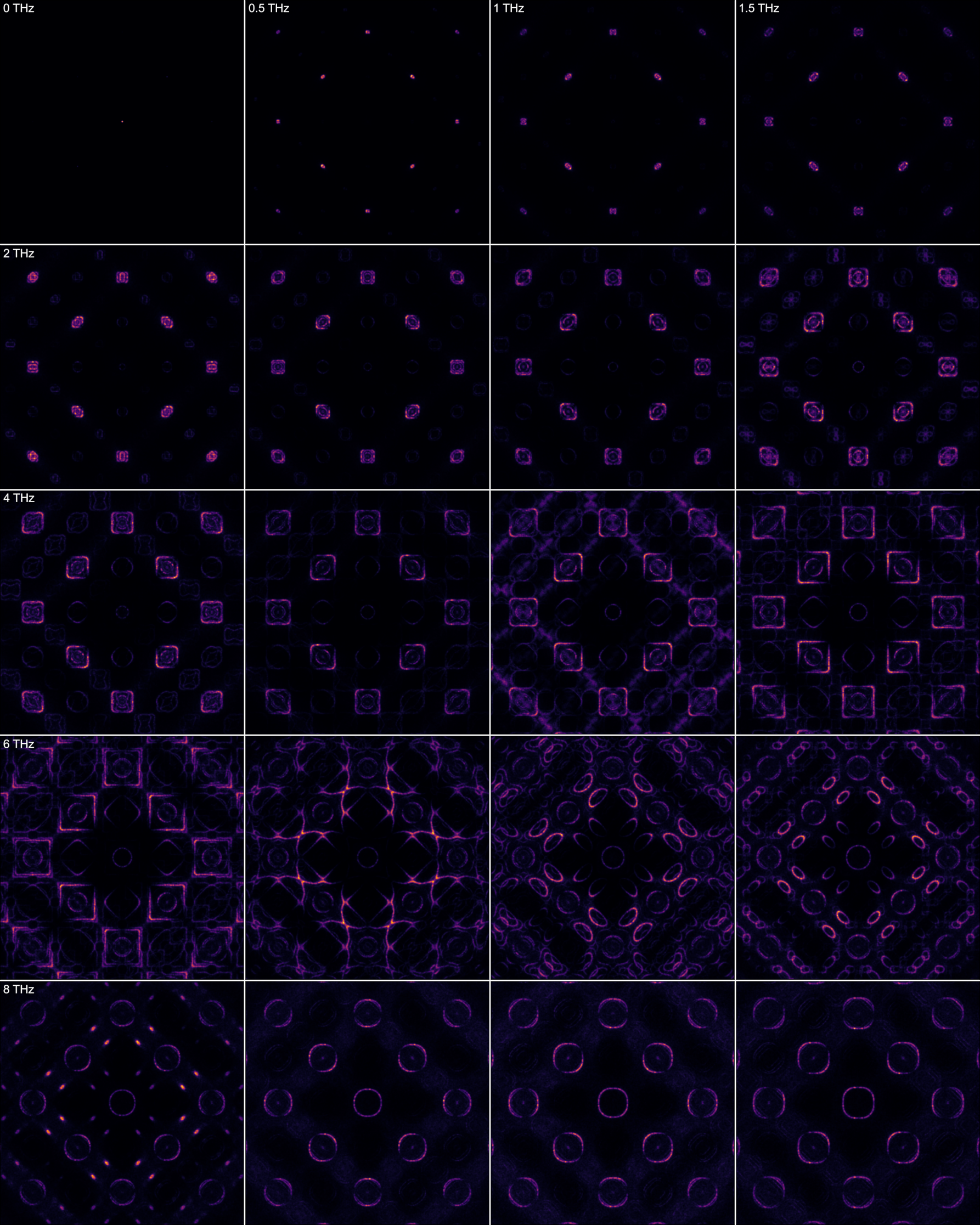}
	\label{fig:ESLICES_A}
\end{figure*}

\begin{figure*}[!tp]
	\caption{}
	\centering
	\includegraphics[width=0.95\linewidth]{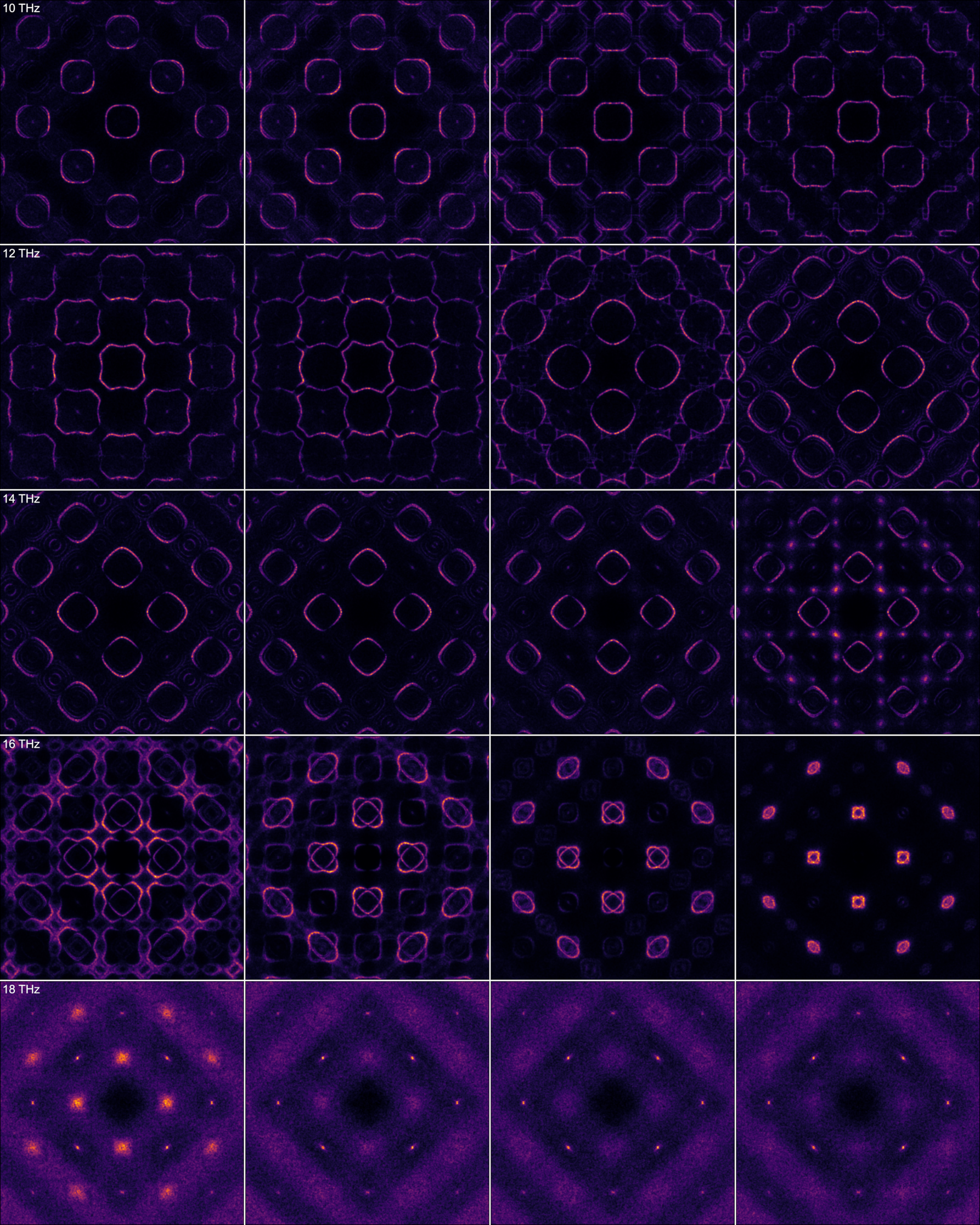}
	\label{fig:ESLICES_B}
\end{figure*}

We have included the momentum-resolved diffraction images (or iso-energy phonon dispersion surfaces) for Stillinger-Weber silicon, at 0.5 THz increments, in Figures \ref{fig:ESLICES_A} and \ref{fig:ESLICES_B}. 

\vspace{120pt}

\subsection{Presence of forbidden reflection in abTEM}
\begin{figure}
    \centering
    \includegraphics[width=0.5\linewidth]{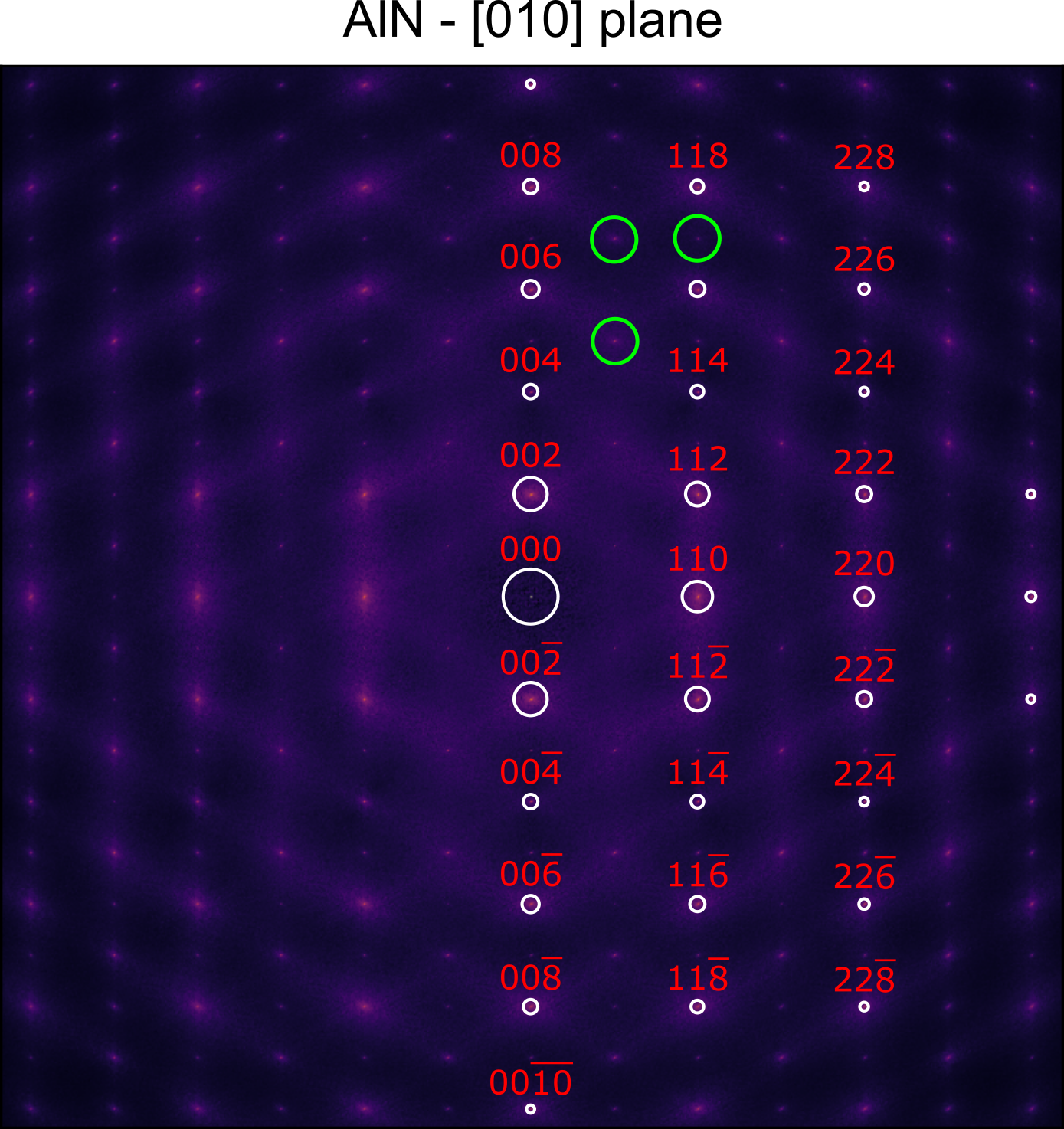}
    \caption{A full diffraction pattern from abEELS (which includes dynamic effects) is shown, with the allowed Bragg reflections (white) as calculated by py4DSTEM. Some forbidden reflection are visible (a subset are circled in green).}
    \label{fig:forbiddenbragg}
\end{figure}

In the main manuscript, we allude to the presence of dynamic scattering effects at high $q$. To support this claim, we compare the diffraction pattern for AlN in the [010] plane between abTEM (which includes dynamic effects) and p4yDSTEM (which captures kinematic effects only). This is shown in Figure \ref{fig:forbiddenbragg}.

\subsection{Noise levels in convergent-beam q-EELS v-DOS}
\begin{figure}
    \centering
    \includegraphics[width=1.0\linewidth]{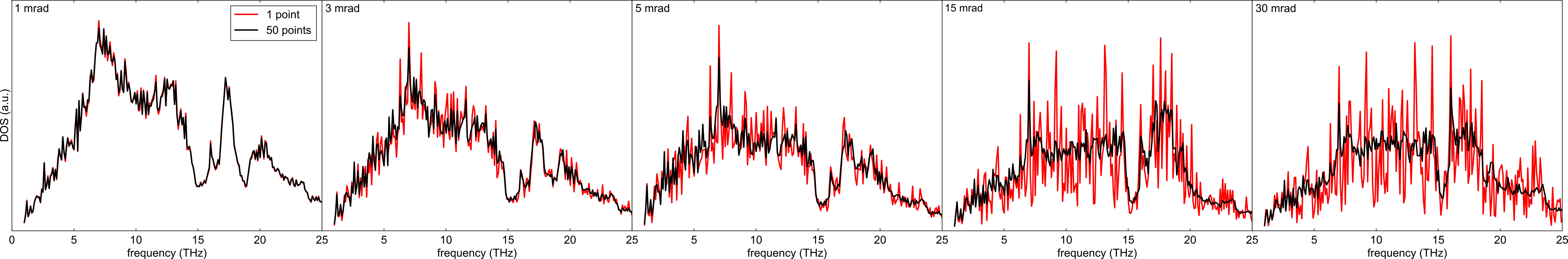}
    \caption{q-EELS was run for 50 probe positions for convergence angles of 1-30 mrad. In the main manuscript, the mean v-DOS across all probe positions was used, shown in black (with an additional low-pass filter for final manuscript figures). the v-DOS for a single probe position is shown in red, and is significantly noisier.}
    \label{fig:abEELSconvergence}
\end{figure}

Two effects yield noisier v-DOS when acquired from a molecular dynamics -based q-EELS simulation. A parallel beam simulation will capture the vibrational behavior over the entire simulation, whereas a convergent beam yields a localized measurement. For a finite-duration simulation, shorter sampling in time, or a smaller region of the system sampled, will result in a noisier signal. this means as the convergence angle is increased, and the signal becomes more localized, the signal will become noisier. The overlapping of Bragg disks, with some level of phase interference between disks, will also result in increased variability between frequencies, multiple MD runs, and between dark-field masks used. For all convergent beam q-EELS v-DOS plots in the main text, q-EELS was run for 50 probe positions, the v-DOS was averaged, and a low-pass filter was applied. In Figure \ref{fig:abEELSconvergence}, we present a comparison of these 50 probe positions (without the low-pass filter), against the v-DOS for a single probe position.

\subsection{Phantom branches within q-EELS phonon dispersions}

\begin{figure*}[t] 
    \centering
    \includegraphics[width=0.95\linewidth]{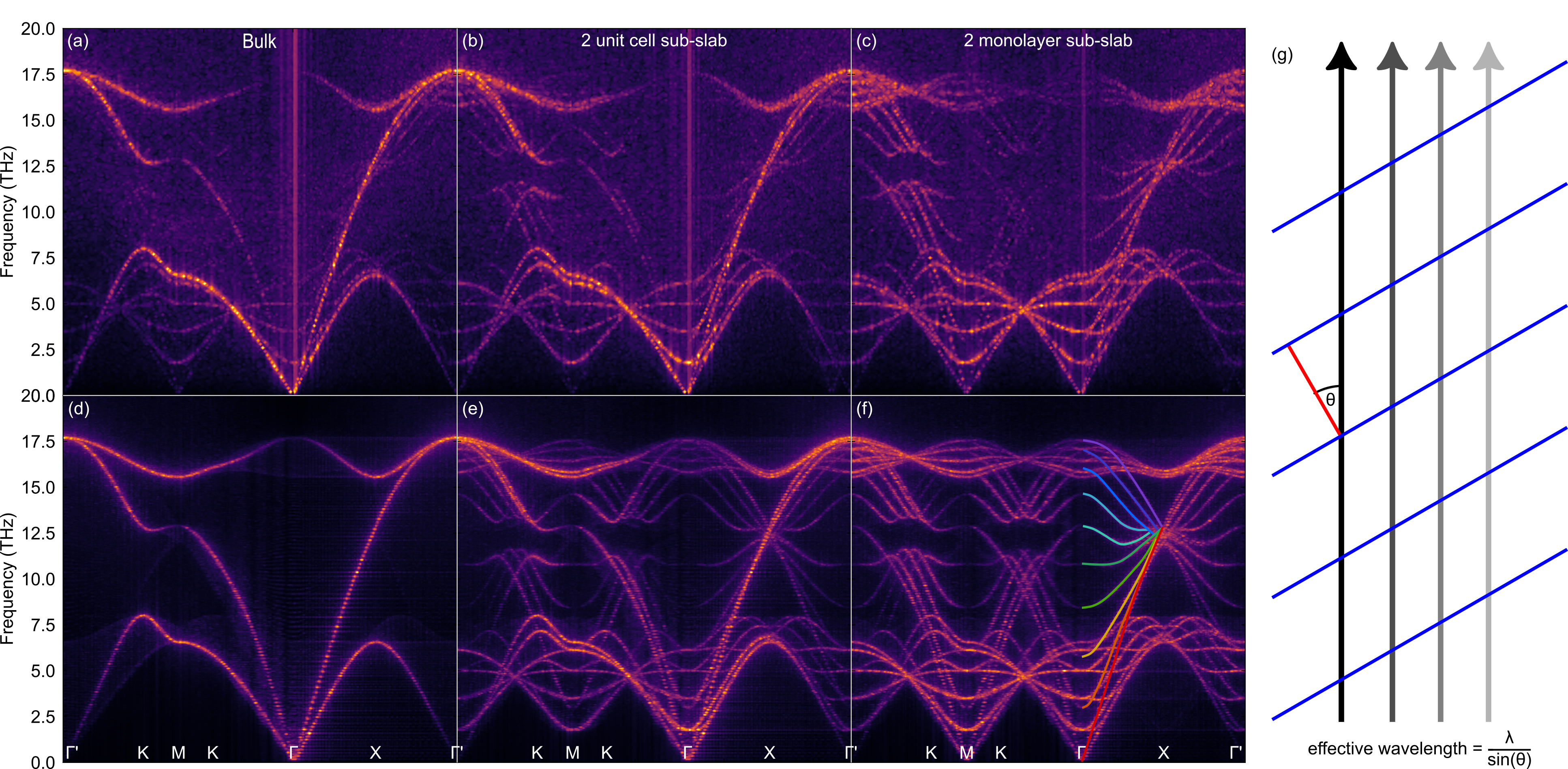}
    \caption{Dispersions are generated along the same reciprocal path for q-EELS (a-c) and coherent probability density, $\protecting{\vv{\bm{\rho}}(\omega,\vv{\bm{k}}})_{coh}$ (d-f) for the bulk (a,d), a slab trimmed to 2 unit cells (b,e), and a slab trimmed to 2 monolayers (c,f). Spurious branches are visible in the bulk q-EELS case, which we attribute to an antenna-like effect, where the projection of through-plane modes can be detected if layers are not sampled uniformly (f). Destructive interference prevents their observation in the bulk calculation, however non-uniform sampling in EELS as a function of depth means they can be observed. In our system, there are 10 discrete out-of-plane branches (highlighted in f), corresponding to the 10 primitive cells in the through-plane direction.}
    \label{fig:monolayers}
\end{figure*}

Spurious branches appear in the simulated q-EELS dispersions in Fig.\ref{fig:PZJACR} and the main manuscript's Fig.2, particularly visible near $\Gamma$. To understand their source, we examine the thickness-dependent results from our q-EELS simulations. We found these phantom branches were brighter for the exit-wave taken from shallower within the simulation, suggesting that these branches are inherent to the system, but that a destructive interference effect is hiding them in the thicker simulations. We also perform the coherent probability density ($\vv{\bm{\rho}}(\omega,\vv{\bm{k}})_{coh}$) calculation across the same structure, by trimming the system to the first several atomic monolayers, and we observe these phantom branches here as well. Dispersions from identical reciprocal paths are shown in Figure \ref{fig:monolayers} for q-EELS (a,b,c) and $\vv{\bm{\rho}}(\omega,\vv{\bm{k}})_{coh}$ (d,e,f), for bulk, a 2 unit cell (8 monolayer), and 2 monolayer trimmed system.

We attribute these phantom branches to an antenna-like effect (schematic shown in Fig. \ref{fig:monolayers}.g). For a wave traveling orthogonal to the sampled direction (blue lines denoting the wavefront), the wave may be picked up with an apparent wavelength corresponding to the sine of the angle between the wave and sampling direction (black or grey arrows), i.e., the projection of the wave onto the sampling direction. Such a feature is not observed in the coherent probability density ($\vv{\bm{\rho}}(\omega,\vv{\bm{k}})_{coh}$) when the entire system is sampled uniformly however (all black and grey arrows weighted equally), as all phases of this orthogonal wave are sampled and interfere destructively. For non-uniform sampling however, full phase cancellation does not occur, and the projected wavelength is observed.

Within q-EELS, the upper layers of the sample are disproportionately sampled. Discrete branches are seen due to a discrete number of through-plane modes, and in the thinnest case, ten branches can be counted (highlighted in \ref{fig:monolayers}.f) corresponding to our MD simulation's ten primitive-cell slab thickness (20 monolayers). For a thick sample, these orthogonal waves will appear as a smearing of the phonon dispersion, as waves exist within the system at all wavelengths and traveling in all directions. For a thin sample however, a small finite number of out-of-plane modes exist. This is the same phenomena which gives us discrete points for modes along $k$ for a finite-sized system.

We used periodic boundary conditions in these simulations, meaning through-plane modes should be expected (waves exiting the top of the slab and re-entering through the bottom), however a subsequent MD simulation with a thin slab in vacuum yielded similar results. For a finite-thickness suspended film, through-plane waves may exist, reflecting off the top and bottom surface. Assuming sufficient measurement sensitivity can be acquired, we predict that these through-plane modes should be observable in experiment. 

\begin{figure}
    \centering
    \includegraphics[width=0.95\linewidth]{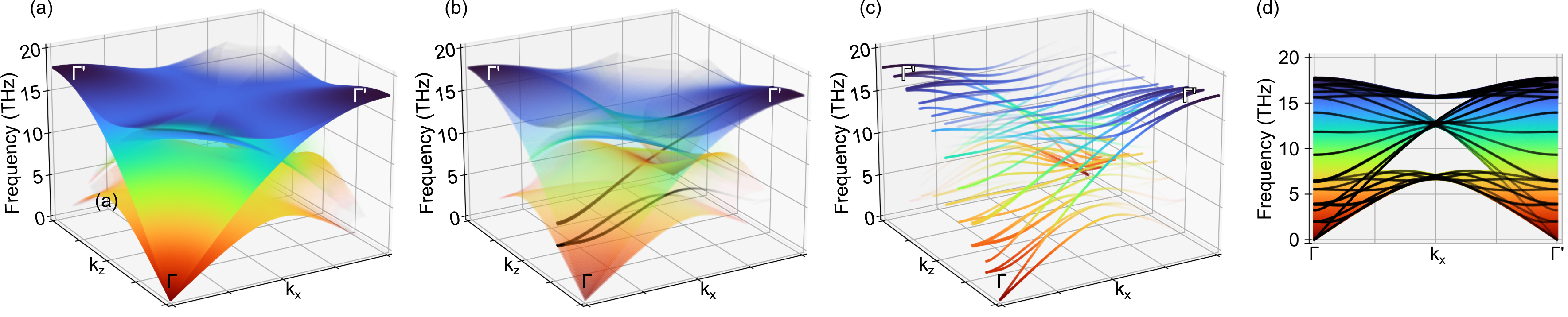}
    \caption{(a) To understand the source of the additional branches seen in the q-EELS simulations, we begin with the 3D phonon surface. (b) For a finite-thickness system, discrete slices across this surface represent the modes present. (c) The number of slices corresponds to the system size; here 10 slices are shown for a slab with a 10 primitive-cell thickness. (d) if these are projected into the $\Gamma$-$X$-$\Gamma'$ direction (via non-uniform sampling of the waves), then the apparent wavevector of each is found, giving the appearance of additional branches.}
    \label{fig:projection}
\end{figure}

To further support our observations of through-plane branches, and to aid in visualization, we also turn to LD. We begin with the 3D phonon dispersion surface, with intensities calculated via coherent summing across the basis, shown in Figure \ref{fig:projection}.a. For a finite-length system, this surface is not continuous and smooth, but rather there are a discrete number of $k$ points along a given direction dependent on the size of the system. For a thin slab, we can approximate the surface as smooth in the semi-infinite direction, but discretized in the thickness direction. A single slice across this surface (at finite $k_z$, across all $k_x$) is shown in Figure \ref{fig:projection}.b. All slices of this surface are shown in Figure \ref{fig:projection}.c. Each point along these curves represents a wave in a low-symmetry direction, i.e., traveling orthogonal to the $x$ direction. If these modes are projected onto $\Gamma$-$X$-$\Gamma'$ (through uneven sampling of these orthogonal waves), the ``side view'' of this surface is found, shown in Figure \ref{fig:projection}.d.

\bibliographystyle{unsrt}
\bibliography{Refs.bib}